\documentclass[a4paper,11pt]{article}
\usepackage[affil-sl,auth-sc]{authblk}
\usepackage{amssymb,amsmath,amsbsy}
\usepackage{mathbbol}
\usepackage{bm}
\usepackage{appendix}
\usepackage{hyperref}
\usepackage[top=1.2in, bottom=1.2in,left=0.9in,includefoot]{geometry}               
\usepackage{cite}
\usepackage{float}
\usepackage{caption}
\usepackage{axodraw}
\usepackage{wasysym}
\usepackage{float}
\usepackage{slashed}
\usepackage{caption}
\usepackage{breqn}
\usepackage{simpler-wick}
\setkeys{breqn}{breakdepth={3}}
\usepackage[utf8]{inputenc}
\newcommand{\nc}{\newcommand}

\newcommand{\bra}[1]{\langle #1|}
\newcommand{\ket}[1]{|#1\rangle}

\newcommand{\ie}{{\it i.e., }}
\newcommand{\eg}{{\it e.g., }}

\newcommand{\bea}{\begin{eqnarray}}
\newcommand{\eea}{\end{eqnarray}}

\nc{\bce}{\begin{center}}     \nc{\ece}{\end{center}}
\nc{\be }{\begin{equation}}   \nc{\ee }{\end{equation}}
\nc{\btb}{\begin{tabular}}    \nc{\etb}{\end{tabular}}
\nc{\f}{\frac}
\nc{\eps}{\varepsilon}
\nc{\vp}{\varphi}

\nc{\tvp}{\widetilde{\varphi}}

\nc{\vpj }{\mbox{${\vp^\dag i\,\raisebox{2mm}{\boldmath ${}^\leftrightarrow$}\hspace{-4mm} D_\mu\,\vp}$}}
\nc{\vpjt}{\mbox{${\vp^\dag i\,\raisebox{2mm}{\boldmath ${}^\leftrightarrow$}\hspace{-4mm} D_\mu^{\,I}\,\vp}$}}

\usepackage[utf8]{inputenc}
\def\eq#1{eq.~(\ref{#1})}

\def\eqs#1#2{eqs.~(\ref{#1}) and (\ref{#2})}
\def\eqss#1#2#3{eqs.~(\ref{#1}), (\ref{#2}) and (\ref{#3})}

\def\eqst#1#2{eqs.~(\ref{#1})--(\ref{#2})}

\DeclareMathOperator{\Tr}{Tr}

\numberwithin{equation}{section}

\DeclareMathOperator{\trp}{\mathsf{T}}
\DeclareMathOperator{\cc}{\mathbb{C}}
\DeclareMathOperator{\gam}{\mathbb{\Gamma}}
\begin{document}

\title{\bf Dirac and Majorana Feynman Rules with four-fermions}
\author{M. Paraskevas\footnote{email: {\tt
      mparask@grads.uoi.gr}}}
\affil{\small Department of Physics, University of Ioannina, \\  GR 45110, Ioannina, Greece}

\date{\today}

\maketitle

\begin{abstract}
%\noindent
A compact method for amplitude calculations in theories with Dirac and Majorana effective operators is discussed. Using the renormalizable formalism of Denner \textit{et al.,}\cite{Denner:1992vza,Denner:1992me} for propagators, vertices and fermion (number) flow and introducing new ``reading-rules", it is shown that fermions can be treated as scalars in the diagrams. The effect of Fermi-statistics appears only in overall signs and is determined once for whole classes of Feynman graphs. Each vertex in this method corresponds to two or more vertices in the standard treatment of effective theories. As such, the advantages of this approach grow together with the number of four-fermion vertices in a given diagram. The discussion develops around effective field theories based on the Standard Model, up to four-fermions and to any order in perturbation theory. Even so, the framework is more general and can be applied elsewhere.
\end{abstract}

\newpage 

\tableofcontents
%\listoffigures
%\listoftables
%%%%%%%%%%%%

\newpage 
\section{Introduction}\label{sec:introduction}

As the experimental accuracy increases, the role of Effective Field Theory (EFT) becomes more and more significant. When applied in the Standard Model (SM) it  can give more accurate predictions for the low-energy observables than the ordinary perturbation theory\cite{Buras:1998raa} (in some cases by far) and also provide information for its unknown UV-completion\cite{Appelquist:1974tg}. Whether one follows the top-down or bottom-up EFT the fundamental formalism remains the same. That is, local operators defined through the Operator Product Expansion (OPE) and typically Feynman diagrams of perturbative-QFT to derive Green's functions. The method presented here is intended for the latter. It can offer a useful tool for easier and consistent derivations of n-point Green's functions involving fermions in EFTs. 

In renormalizable theories fermions appear in the interaction Lagrangian as bilinears, due to Lorentz invariance and power counting rules. Diagrammatically this property appears in the vertex legs which display two fermions and a single boson, scalar or gauge. In effective theories fermion vertices can have more legs. For diagrams involving vertices with only two fermions and any number of bosons, the standard diagrammatic treatment can apply. But this is not the only case. 

Fermions in EFTs can also appear  as products of bilinears. Due to Fermi statistics they form anticommuting tensors in the Dyson expansion. Diagrammatically, anticommutation appears with the usual external-leg permutation and fermion-loop minus signs, and also with new signs due to \textit{internal-leg} permutation. This novel effect is deeply correlated with the other two, making the generalization of the renormalizable treatment very difficult. As usually happens in such cases, one resorts to Wick contractions using the initial Lagrangian coupling or at best some partially symmetrized equivalent form. However, this tedious procedure can be avoided. In the method presented here, the standard vertex setup of renormalizable theories together with the well-known formalism of Denner \textit{et al.,}\cite{Denner:1992vza,Denner:1992me} (to include also fermion-number violating effects) are  sufficient. Modifications appear in the rules for reading the diagrams, in trivial scalar-type combinatorics, and in a more cautious treatment of vertices when attaching legs together to form graphs. But in the end, fermions  in the diagrams are treated as scalars with any remaining effect of Fermi-statistics appearing as overall sign in graphs. This sign is determined again through Wick-contractions but here once for whole classes of diagrams. 
  
The work of the Warsaw U., group\cite{Grzadkowski:2010es} together with the development of computer-codes for automatized calculations \cite{Aebischer:2017ugx,Brivio:2017btx,Celis:2017hod,Falkowski:2015wza} have given a significant boost of interest to the bottom-up EFT approach\cite{Alonso:2013hga,Dedes:2017zog,Brivio:2017vri}. This has driven together other EFTs based on Standard Model (SM) field content \cite{Brivio:2016fzo} or minimal extensions of it\cite{Alanne:2017oqj,Crivellin:2016ihg}. It has also given a renewed interest in well-known EFT frameworks\cite{Jenkins:2017dyc,Jenkins:2017jig}, extensively used in the past with remarkable accuracy. To avoid being vague and keep up with the recent developments in SM-based EFTs the discussion evolves around them. Some of the examples given in main text and appendix are taken from the Feynman Rules of $R_\xi$-SMEFT\cite{Dedes:2017zog}. However, for the purpose of a more clear illustration on the main points of this method, they were simplified and adapted accordingly\footnote{The Feynman Rules of $R_\xi$-SMEFT, as found in later versions of Ref.~\cite{Dedes:2017zog}, are consistent with the formalism presented here.}.

%%%%%%%%%%%%%%%%%%%%%%%%%%%%%%%%%
\section{Basic Formalism}\label{sec:basic}
The method proposed is based on the formalism and conventions of \cite{Denner:1992vza,Denner:1992me} which were developed for renormalizable theories. It is useful to review them and focus on some key-points and aspects of the treatment. In many cases explicit indices will be used in the relations which are no different from those in the references. This extended notation will play an important role in promoting the (standard) matrix treatment of renormalizable theories into a tensor treatment for the general case. 

The conventions set here are more restrictive than those in Ref.\cite{Denner:1992vza}. In fact, they are adapted to follow Peskin\&Schroeder \cite{Peskin:1995ev} so that the reader can directly refer to the textbook for spin sums and various other useful identities. One can always switch to some other set of conventions as long as the more relaxed definitions and derived relations of \cite{Denner:1992vza}, still hold.

\subsection{Notation and conventions}
In addition to the notation introduced in Fig.\ref{fig:not}  
\begin{figure}[b] 
\begin{center}
\begin{tabular}{lp{1cm}l}
\hline
Index type  & & Symbols \\
\hline
Tensor index (see text) && $i=1,2,\dots$ \\
Flavor (generation) && $f_i$ \\
Spinor && $s_i$ \\
Spin && $\bar s_i$ \\
Color or other && $m_i$ \\
Lorentz && $\mu$ \\
\hline
\end{tabular}
\end{center}
\caption{Notation for explicit indices used throughout the paper. \textit{Primed indices} are always assumed contracted (dummy). \textit{Spinorial indices} are always contracted in amplitudes even if primed indices are not used for them.}
\label{fig:not}
\end{figure}
it will be useful to assign number labels for sets of indices as, 
$$i \equiv \{ f_i, s_i, m_i \}.$$
For index contractions, it is assumed that only relevant indices in these sets are summed over. For example, creation/annihilation (ladder) operators do not carry spinorial indices and the same obviously stands for lepton fields with respect to color. 

Although not necessary, choosing some specific basis for Dirac-matrices  simplifies considerably manipulations. In the \textit{chiral} representation,
\begin{align*}
\gamma^{\mu} = \left(\begin{array}{cc}
\mathbb{0} & \sigma^{\mu} \\ 
\bar{\sigma}^{\mu} & \mathbb{0}
\end{array}\right)  ~,~~\sigma^\mu = (\mathbb{1}, \sigma^i)~,~~ \bar{\sigma}^\mu = (\mathbb{1},- \sigma^i)~,
\end{align*}
with $\sigma^i$ being the Pauli matrices and Minkowski signature $\eta_{\mu\nu}= \text{diag}(+,-,-,-)$ . The quantized Dirac field and its conjugate read,
\begin{align}
\psi(x) &= \int {d^3k\over (2\pi)^3 \sqrt{2E}}\sum_{\bar s} \Big[b(k,\bar{s}) u(k,\bar s) e^{-ikx} + d^{\dag} (k,\bar s) v(k,\bar s) e^{ikx} \Big]~,\nonumber\\
\bar\psi(x) &= \int {d^3k\over (2\pi)^3 \sqrt{2E}}\sum_{\bar s} \Big[b^{\dag}(k,\bar s) \bar u(k,\bar s) e^{ikx} + d(k,\bar s) \bar v(k,\bar s) e^{-ikx} \Big]~,\label{psiquant}
\end{align}
with the creation (annihilation) operators $b^\dag(b)$ and $d^\dag(d)$ referring to fermions and antifermions, respectively. 
 
Fermion-number violating effects are also included in the formalism by introducing \emph{charge-conjugate (cc)} fields and operators. A suitable choice for the unitary cc-matrix in the chiral basis is,
\begin{align} 
&\cc \equiv -i \gamma^2 \gamma^0 ~,\\
\cc^\dag = &\cc^{-1}= \cc^{\trp} = -\cc ~. 
\end{align}
One can then relate the Dirac spinors and introduce cc-fields, through
\begin{align}
u(p,\bar s)=\cc\bar v^\trp (p,\bar s)~~&,~~~  v(p,\bar s)=\cc\bar u^\trp (p,\bar s)~,\label{ccspinors}\\ 
\psi^c = \cc\bar \psi^\trp ~~&,~~~  \bar\psi^c=\psi^\trp \mathbb{C}~.  \label{psicc}
\end{align} 
Majorana fermions are defined to satisfy in addition the Majorana condition in the form, 
\bea 
\nu = \nu^c=\cc\bar \nu^\trp ~,
\eea
as is the case for neutrinos in effective field theories based on SM. 
For Majorana fields there is no distinction between particles and antiparticles. The quantized fields and their conjugates can be trivially obtained from \eq{psiquant}, for $d=b$. The Dirac cc-fields  are also obtained from \eq{psiquant} by applying \eqs{ccspinors}{psicc}. The analytical expressions are equivalent to those given in Ref.\cite{Denner:1992vza}, up to different normalizations. 

It is also useful to display for later reference a crucial property of the $\cc$-matrix related to fermion-number violating diagrams. The charge conjugate of some Dirac  matrix structure defined as,
\begin{align}
\Gamma^c \equiv \cc \Gamma^\trp \cc^{-1}~, \label{gammac}
\end{align}
can be further simplified with the relation (no sum over $i$),
\begin{align}
&\quad \cc \Gamma_i^\trp \cc^{-1}=\eta_i \Gamma_i~,  \nonumber
\\
\eta_i &= \left\{\begin{array}{l} 
+ 1  \textnormal{ for } \Gamma_i = 1, i\gamma_5, \gamma_\mu \gamma_5\\
-1 \textnormal{ for } \Gamma_i = \gamma_\mu, \sigma_{\mu\nu}~.
\end{array}\right. \label{vertgammac} 
\end{align}
Note in particular that if $\Gamma$ is a Dirac matrix structure without $\cc$-dependence, neither $\cc$ nor transpose matrices will appear in $\Gamma^c$.

\subsection{Propagators and external fermions}
Propagators and external Dirac spinors appearing in QFT calculations can be associated with Wick contractions of fermionic operators. The formal correspondence, starting from time ordered products in spacetime and taking Fourier transformations is not repeated here. Raising the level of mathematical rigor would obscure the main points of the treatment and in particular its generalization, later on. 

Considering momentum $p$ flowing always from $1$ to $2$, the Dirac and Majorana propagators read, 
\begin{align}
\wick{\c{\psi_{2}}\c{\bar \psi_{1}}} &\to  \Big({i\over \slashed{p}-m_{f_1}}\Big)_{s_2 s_1}\delta_{f_1 f_2}\delta_{m_1m_2}\equiv i S_{21}(p) ~,\nonumber\\
&\begin{array}{c}
\begin{picture}(80,30)
\ArrowLine(0,10)(80,10)
\Text(0,0)[r]{$1$}
\Text(80,0)[l]{$2$}
\Text(40,25)[c]{$\psi,\nu$}
\end{picture}
\end{array}  
~~~\to ~i S_{21}(p) ~.\label{S}
\end{align}
For the charge-conjugate fields, one obtains,
\begin{align}
\wick{\c{\psi^c_{2}}\c{\bar \psi^c_{1}}} \to  \Big[\cc\Big({i\over \slashed{p}-m_{f_1}}\Big)^\trp\cc^{-1}\Big]_{s_2 s_1}\delta_{f_1 f_2}\delta_{m_1m_2}\equiv  i S^c_{21}(p)& ~,\nonumber\\
\begin{array}{c}
\begin{picture}(80,30)
\ArrowLine(0,10)(80,10)
\Text(0,0)[r]{$1$}
\Text(80,0)[l]{$2$}
\Text(40,25)[c]{$\psi^c$}
\end{picture}
\end{array}  
~~~\to ~ i S^c_{21}(p) = i S_{21}(-p)~,~~~~~&\label{Sc}
\end{align}
so that also this propagator is $\cc$-independent. Note that only these propagators are required for amplitude calculations within this formalism.

For processes involving external fermion states, Dirac four-component (commuting) spinors are induced through field-ladder Wick contractions. Showing momentum arrows on top, they read,
\begin{align}
\wick{\c{\psi_{1}}\c{b}^\dag_{1}}\to u_{1}(p_1,\bar s_1) :~~~~~\begin{array}{c}
\begin{picture}(40,25)
\ArrowLine(0,10)(40,10)
\Line(10,15)(20,15)
\ArrowLine(19,15)(20,15)
\CCirc(40,10){3}{Black}{Black}
\Text(-7,10)[c]{$\psi_{1}$}
\end{picture}
\end{array}  
~~~
&~,~~~~~&~~~~ \wick{\c{b}_{1}\c{\bar \psi_{1}}}\to \bar u_{1}(p_1,\bar s_1): ~~~~ \begin{array}{c}
\begin{picture}(40,25)
\ArrowLine(40,10)(0,10)
\Line(30,15)(20,15)
\ArrowLine(20,15)(19,15)
\CCirc(40,10){3}{Black}{Black}
\Text(-7,10)[c]{$\psi_{1}$}
\end{picture}
\end{array}  
~~~,&
\nonumber\\[-5mm]
\wick{\c{\bar\psi_{1}}\c{d}^\dag_{1}}\to \bar v_{1}(p_1,\bar s_1) : ~~~~~\begin{array}{c}
\begin{picture}(40,25)
\ArrowLine(40,10)(0,10)
\Line(10,15)(20,15)
\ArrowLine(19,15)(20,15)
\CCirc(40,10){3}{Black}{Black}
\Text(-7,10)[c]{$\psi_{1}$}
\end{picture}
\end{array}  
~~~
&~,~~~&~~~~
\wick{\c{d}_{1}\c{\psi_{1}}}\to
v_{1}(p_1,\bar s_1):~~~~\begin{array}{c}
\begin{picture}(40,25)
\ArrowLine(0,10)(40,10)
\Line(30,15)(20,15)
\ArrowLine(20,15)(19,15)
\CCirc(40,10){3}{Black}{Black}
\Text(-7,10)[c]{$\psi_{1}$}
\end{picture}
\end{array} ~~~,& \label{psiext} 
\end{align}
\begin{align}
\wick{\c{\psi^c_{1}}\c{d}^\dag_{1}}\to u_{1}(p_1,\bar s_1) :~~~~~\begin{array}{c}
\begin{picture}(40,25)
\ArrowLine(0,10)(40,10)
\Line(10,15)(20,15)
\ArrowLine(19,15)(20,15)
\CCirc(40,10){3}{Black}{Black}
\Text(-7,10)[c]{$\psi^c_{1}$}
\end{picture}
\end{array}  
~~~
&~,~~~~~&~~~~ \wick{\c{d}_{1}\c{\bar \psi^c_{1}}}\to \bar u_{1}(p_1,\bar s_1): ~~~~\begin{array}{c}
\begin{picture}(40,25)
\ArrowLine(40,10)(0,10)
\Line(30,15)(20,15)
\ArrowLine(20,15)(19,15)
\CCirc(40,10){3}{Black}{Black}
\Text(-7,10)[c]{$\psi^c_{1}$}
\end{picture}
\end{array}  
~~~,~&  
\nonumber\\[-5mm]
\wick{\c{\bar\psi^c_{1}}\c{b}^\dag_{1}}\to \bar v_{1}(p_1,\bar s_1) :~~~~~\begin{array}{c}
\begin{picture}(40,25)
\ArrowLine(40,10)(0,10)
\Line(10,15)(20,15)
\ArrowLine(19,15)(20,15)
\CCirc(40,10){3}{Black}{Black}
\Text(-7,10)[c]{$\psi^c_{1}$}
\end{picture}
\end{array}  
~~~
&~,~~~~~&~~~~
\wick{\c{b}_{1}\c{\psi^c_{1}}}\to
v_{1}(p_1,\bar s_1): ~~~~\begin{array}{c}
\begin{picture}(40,25)
\ArrowLine(0,10)(40,10)
\Line(30,15)(20,15)
\ArrowLine(20,15)(19,15)
\CCirc(40,10){3}{Black}{Black}
\Text(-7,10)[c]{$\psi^c_{1}$}
\end{picture}
\end{array} ~~~,~& \label{psiextc}  
\end{align}
while for Majorana spinors ($\nu=\nu^c$) they can be easily derived from any of the above, for $d=b$. 

In this notation arrows on fermion lines always denote \textit{fermion flow}.  Contrary to  Ref. \cite{Denner:1992vza} fermion-number arrows are suppressed. Instead, the cc-fields are explicitly shown.  In case needed, note that for the ordinary Dirac fields fermion-number and fermion arrows have the same direction. For cc-fields they have opposite direction while for Majorana fields, fermion-number is meaningless.

\subsection{Fermion flow and the first steps for an EFT generalization}
\label{sec:fflow}
The fermion flow is a generalization of the well known fermion-number flow which applies universally on all diagrams. For  fermion-number conserving graphs the two are identical. Notice the Dirac and Majorana propagators, arrows and external spinors,  which are the usual ones. The distinction happens only for fermion-number violating processes. There, one still proceeds diagrammatically as usual, by reading the diagram opposite to the arrow direction\footnote{This ``reading-rule" will be dropped in the EFT generalization.}. Using the cc-fields and when needed the cc-vertices, clashing arrows never appear in diagrams. The standard diagrammatic picture of propagating fermionic arrows in open chains or closed loops is preserved, the three propagators discussed previously are the only ones explicitly used, while $\cc$ and transpose $\gamma$-matrices are never present in the final results.  Some key-points of this treatment and the first steps for a consistent generalization in effective theories are now discussed. More details for the renormalizable case can be found in the references.

The $\cc$-matrix in the couplings of fermion-number violating operators can be absorbed by introducing cc-fields as in \eq{psicc}. All Lagrangian couplings therefore become $\cc$-independent and fermions appear through field-conjugate bilinears. The same holds for effective theories with the only difference that products of bilinears are also allowed. A relevant term in the Lagrangian can have the form,
\begin{align*}
\mathcal{L}_{int} =  \bar \xi_{1'} \gam_{1'2'} \chi_{2'}^c \Phi~,
\end{align*}
where $\chi,\,\xi$ are some fermion fields and $\Phi$ is a single boson field in renormalizable theories or any bosonic operator (\ie field or product of fields) for effective theories. In  principle, $\Phi$ can involve fermion bilinears but for the moment this case is ignored. It can also include flavor or other quantum numbers but for the analysis here it makes no difference. What only matters is that boson operators commute with everything else and spinorial indices for $\Phi$ are irrelevant. Under these considerations and without loss of generality the Lagrangian coupling can have the form,
\begin{align*}
\mathbb{\Gamma}_{12}=C_{f_1f_2} \Gamma_{s_1s_2}~,
\end{align*}
where as usual $\Gamma$ is some arbitrary Dirac matrix structure independent of $\cc$, as explained.

By definition, cc-fields are not independent and therefore the  propagators/contractions, 
$$\wick{(\c{\psi}\c{\psi^c}),(\c{\bar\psi^c}\c{\bar\psi}), (\c{\nu}\c{\nu}), (\c{\bar\nu} \c{\bar\nu}) },$$
are non-vanishing. Nevertheless, they can be rearranged with \eq{psicc}, as,  
$$\wick{\c1{\chi_1}\dots (\c2{\bar\xi_{1'}} i\mathbb{\Gamma}_{1'2'}  \c1{\chi^c_{2'}}) \dots \c2{\bar\xi^c_{2}}}= \wick{\c1{\chi_1} \dots (\c1{\bar\chi_{1'}} i\mathbb{\Gamma}^c_{1'2'}  \c1{\xi^c_{2'}}) \dots \c1{\bar\xi^c_{2}}} ~,$$
with, 
$$\mathbb{\Gamma}^c_{12}=(\cc \mathbb{\Gamma}^\trp \cc^{-1})_{1 2} = C_{f_2f_1} \Gamma^c_{s_1s_2}~,$$
being the cc-coupling. Thus, only the ordinary propagators/contractions and $\Gamma^c$ appear after the rearrangement, with all of them being $\cc$-independent according to \eqss{gammac}{vertgammac}{Sc}. When color or other quantum numbers are relevant, transposition with respect to those indices needs to be considered, as well. 

In diagrammatic representation the rearrangement is equivalent to,
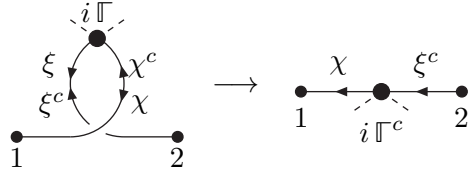
\begin{align*}
~~~~~~~\begin{picture}(40,40)
\DashLine(0,20)(10,27){3}
\Text(0,30)[c]{$i\gam$}
\DashLine(0,20)(-10,27){3}
\ArrowArc(-10,3)(20,13,14)
\ArrowArcn(-10,3)(20,-13,-14)
\ArrowArc(10,3)(20,169,170)
\ArrowArcn(10,3)(20,-169,-170)
\CArc(10,3)(20,120,230)
\CArc(10,3)(20,250,270)
\CArc(-10,3)(20,-90,60)
\Line(30,-17)(10.,-17)
\Line(-30,-17)(-10.,-17)
\CCirc(0,20){3}{Black}{Black}
\Text(-17,10)[c]{$\xi\,$}
\Text(-17,-5)[c]{$\xi^c$}
\Text(17,-5)[c]{$\chi\,$}
\Text(17,10)[c]{$\chi^c$}
\Text(-30,-25)[c]{$1$}
\CCirc(-30,-17){2}{Black}{Black}
\CCirc(30,-17){2}{Black}{Black}
\Text(30,-25)[c]{$2$}
\end{picture} 
\longrightarrow\hspace{1.5cm}
\begin{picture}(40,40)
\DashLine(0,0)(10,-7){3}
\Text(0,-15)[c]{$i\gam^c$}
\DashLine(0,0)(-10,-7){3}
\ArrowLine(0,0)(-30,0)
\ArrowLine(30,0)(0,0)
\CCirc(0,0){3}{Black}{Black}
\Text(-15,10)[c]{$\chi\,$}
\Text(15,10)[c]{$~\xi^c$}
\Text(-30,-10)[c]{$1$}
\CCirc(-30,0){2}{Black}{Black}
\CCirc(30,0){2}{Black}{Black}
\Text(30,-10)[c]{$2$}
\end{picture}\\[1cm]
\end{align*}
with dashed lines denoting the presence of $\Phi$, which is obviously irrelevant. As should be clear, the introduction of cc-vertices eliminates clashing arrows from diagrams and, equivalently, the unwanted propagators from the analytic expressions. For $\chi=\xi$ one has $\mathbb{\Gamma}=\mathbb{\Gamma}^c$ and for Majorana fermions ($\nu=\nu^c$) in addition,  
\begin{align*}
\wick{\c1{\nu_1}\dots  (\c2{\bar\nu_{1'}} i\mathbb{\Gamma}_{1'2'}  \c1{\nu_{2'}}) \dots \c2{\bar\nu_{2}}}= \wick{\c1{\nu_1}\dots  (\c1{\bar\nu_{1'}} i\mathbb{\Gamma}_{1'2'}  \c1{\nu_{2'}})\dots  \c1{\bar\nu_2}}~,
\\
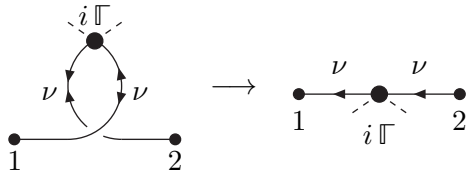
\begin{picture}(40,40)
\DashLine(0,20)(10,27){3}
\Text(0,30)[c]{$i\gam$}
\DashLine(0,20)(-10,27){3}
\ArrowArc(-10,3)(20,13,14)
\ArrowArcn(-10,3)(20,-13,-14)
\ArrowArc(10,3)(20,169,170)
\ArrowArcn(10,3)(20,-169,-170)
\CArc(10,3)(20,120,230)
\CArc(10,3)(20,250,270)
\CArc(-10,3)(20,-90,60)
\Line(30,-17)(10.,-17)
\Line(-30,-17)(-10.,-17)
\CCirc(0,20){3}{Black}{Black}
\Text(-17,0)[c]{$\nu$}
\Text(17,0)[c]{$\nu$}
\Text(-30,-25)[c]{$1$}
\CCirc(-30,-17){2}{Black}{Black}
\CCirc(30,-17){2}{Black}{Black}
\Text(30,-25)[c]{$2$}
\end{picture} 
\longrightarrow\hspace{1.5cm}
\begin{picture}(40,40)
\DashLine(0,0)(10,-7){3}
\Text(0,-15)[c]{$i\gam$}
\DashLine(0,0)(-10,-7){3}
\ArrowLine(0,0)(-30,0)
\ArrowLine(30,0)(0,0)
\CCirc(0,0){3}{Black}{Black}
\Text(-15,10)[c]{$\nu$}
\Text(15,10)[c]{$\nu$}
\Text(-30,-10)[c]{$1$}
\CCirc(-30,0){2}{Black}{Black}
\CCirc(30,0){2}{Black}{Black}
\Text(30,-10)[c]{$2$}
\end{picture}~~~~~~~~~\\[0.5cm]
\end{align*}
so that the unwanted contractions for Majorana fields are always equivalent to the ordinary ones without any additional consideration.

The unwanted propagators denote the explicit presence of $\cc$ in the theory. Nevertheless, with successive cc-operations as the ones shown previously, one can directly absorb the $\cc$-matrix  or send it to the external legs. This is a general property irrespective of diagram or number of contractions. In the legs, it can be eventually eliminated through the spinor or field redefinitions of \eq{ccspinors} or \eq{psicc}. In diagrammatic language (notice arrows),

$$~~~~~~~~~~~~~~~~~~~\begin{picture}(40,40)
\Text(-17,5)[c]{$\xi\,$}
\Text(16,-3)[c]{$\chi$}
\Text(17,10)[c]{$\,\chi^c$}
\DashLine(0,20)(7,27){3}
\DashLine(0,20)(-8,27){3}
\Text(0,34)[c]{$i\gam_{(b)}$}
\DashLine(15,-17)(21,-24){3}
\DashLine(15,-17)(9,-24){3}
\Text(15,-34)[c]{$i\gam_{(a)}$}
\DashLine(-15,-17)(-21,-24){3}
\DashLine(-15,-17)(-9,-24){3}
\Text(-15,-34)[c]{$i\gam_{(c)}$}
\ArrowArc(-10,3)(20,13,14)
\ArrowArcn(-10,3)(20,-13,-14)
\ArrowArc(10,3)(20,179,180)
\CArc(10,3)(20,120,230)
\CArc(10,3)(20,250,280)
\CArc(-10,3)(20,260,60)
\CCirc(0,20){3}{Black}{Black}
\CCirc(-15,-17){2}{Black}{Black}
\CCirc(15,-17){2}{Black}{Black}
\ArrowLine(15,-17)(35,-17)

\DashArrowLine(-15,-17)(-40,-17){1}
\Text(45,-17)[c]{$\psi_2$}
\Line(20,-12)(25,-12)
\ArrowLine(25,-12)(26,-12)
\end{picture}~~\sim \dots \bar u_2(p_2,\bar s_2)\hspace{0.5cm}
\longrightarrow
\hspace{2cm}
\begin{picture}(40,40)
\Text(-6,-7)[c]{$\chi$}
\Text(36,-3)[c]{$~\,\xi^c$}
\Text(37,10)[c]{$\xi$}
\DashLine(20,20)(27,27){3}
\DashLine(20,20)(12,27){3}
\Text(20,34)[c]{$i\gam_{(a)}$}
\DashLine(5,-17)(10,-24){3}
\DashLine(5,-17)(-1,-24){3}
\Text(5,-34)[c]{$i\gam_{(b)}^c$}
\DashLine(-20,-17)(-26,-24){3}
\DashLine(-20,-17)(-14,-24){3}
\Text(-20,-34)[c]{$i\gam_{(c)}$}
\ArrowArc(10,3)(20,13,14)
\ArrowArcn(10,3)(20,-13,-14)
\ArrowArc(30,3)(20,179,180)
\CArc(30,3)(20,120,230)
\CArc(30,3)(20,250,270)
\CArc(10,3)(20,260,60)
\CCirc(20,20){3}{Black}{Black}
\CCirc(-20,-17){2}{Black}{Black}
\CCirc(5,-17){2}{Black}{Black}
\Text(58,-17)[c]{$\psi_2$}
\Line(48,-17)(30,-17)
\ArrowLine(5,-17)(-20,-17)
\DashArrowLine(-20,-17)(-45,-17){1}
\Line(35,-12)(40,-12)
\ArrowLine(40,-12)(41,-12)
\end{picture}~~~~~~~~~~~~~~\sim \dots \bar u_2(p_2,\bar s_2)\hspace{0.6cm}
$$\\[1cm]
$$
\hspace{6.7cm}\longrightarrow
\hspace{2.cm}
\begin{picture}(40,40)
\Text(-6,10)[c]{$\chi$}
\Text(20,10)[c]{$\xi^c$}
\DashLine(30,0)(36,-6){3}
\DashLine(30,0)(24,-6){3}
\Text(30,-17)[c]{$i\gam_{(a)}^c$}
\DashLine(5,0)(10,-6){3}
\DashLine(5,0)(-1,-6){3}
\Text(5,-17)[c]{$i\gam_{(b)}^c$}
\DashLine(-20,0)(-26,-6){3}
\DashLine(-20,0)(-14,-6){3}
\Text(-20,-17)[c]{$i\gam_{(c)}$}
\CCirc(30,0){2}{Black}{Black}
\CCirc(-20,0){2}{Black}{Black}
\CCirc(5,0){2}{Black}{Black}
\Text(60,0)[c]{$\psi_2^c$}
\ArrowLine(50,0)(30,0)
\ArrowLine(30,0)(5,0)
\ArrowLine(5,0)(-20,0)
\DashArrowLine(-20,0)(-45,0){1}
\Line(35,5)(40,5)
\ArrowLine(40,5)(41,5)
\end{picture}\hspace{1.2cm}~~~\sim \dots v_2(p_2,\bar s_2)~~~~~~~~~~~~~~~~~~~
$$\\[1cm]
with the subscripts in $i\gam$'s implying that different vertices may be involved, in general.
 
From the practical side, one never needs to worry about the unwanted propagators or such tedious manipulations. For fermion-number violation with Dirac fields, one considers the cc-fields/propagators and when needed the cc-vertices\footnote{The cc-operation applies on bilinears and therefore four-fermions or higher operators can produce more than one cc-vertex. Only the vertex which preserves the arrow flow in a given diagram will be needed also in EFT.} to avoid clashing arrows. That is, going directly to the last diagram above. For Majorana fields the cc-operation is trivial. In both cases, considering  the diagrams which preserve the arrow flow will suffice also in the EFT generalization\footnote{Few exceptions referring to the four-Majorana vertex are discussed separately in Appendix \ref{app:Remarks}.}. Any other effect of such contractions (which are always present in the theory) will be already included through trivial scalar-type combinatorics for fermion diagrams. 

\section{Generalization to EFT}\label{sec:generalization}
Unlike propagators,  the vertices of a theory are not unique and one can define them in various ways. In fact, they are not even required for amplitude calculations since all Green's functions can be derived alternatively from the Lagrangian through first principles. That is, Wick contractions on the Dyson expansion, a method which will be extensively used here to extract vertices (as usual), verify derivations and resolve diagrammatic ambiguities. 

Even though not essential, the vertex is one of the most valuable shortcuts for amplitude calculations. A suitable and \textit{consistent} choice for it can reduce substantially the complexity in calculations and thus the possibility of mistakes. The four-fermion vertices here will be derived from \textit{matrix elements}. This is the standard way vertices are defined in renormalizable theories.   
\subsection{The vertex setup}\label{sec:vert}
A suitable and general enough four-fermion Lagrangian can have the form,  
\begin{align}
\mathcal{L}_{int} \equiv \sum C^{f_1f_2f_3f_4} \Gamma_{s_1 s_2}\Gamma_{s_3 s_4}~ \bar\psi^{f_1}_{s_1} \psi^{f_2}_{s_2 } \bar\psi^{f_3}_{s_3 } \psi^{f_4}_{s_4}~, \label{Lpsi}
\end{align}
with $\psi$ being a Dirac-fermion of a certain type\footnote{For different type the same discussion applies in a more trivial manner.} (\eg charged lepton, $e$), 
 $C$ being a tensor coupling with flavor indices (\eg Wilson coefficient), $\Gamma$ some arbitrary Dirac matrix structure (\eg $\gamma_\mu P_L $) and the summation referring to all indices. The two $\Gamma$'s can be equal or different and makes no difference for the setup. Obviously it  matters for the vertex expression and to keep things simple the former is chosen. The flavor coupling is then  symmetric under the interchange of fermion bilinears in \eq{Lpsi}, satisfying, 
\bea 
C^{(12)(34)} = C^{(34)(12)}~.\label{Csym}
\eea  
Other couplings and indices referring to color or else follow the treatment of the flavor coupling and therefore can be suppressed.   

The vertex is defined as for renormalizable theories. Considering a momentum convention (\ie here all incoming), a reference order for ladder operators and taking all possible contractions with the interaction Hamiltonian ($\mathcal{H}_{int}=-\mathcal{L}_{int}$) in the first order of the Dyson expansion, one obtains, 
\begin{align}
\bra{0} (-i \mathcal{H}_{int}) b^\dag_4 d^\dag_3 b^\dag_2 d^\dag_1 \ket{0}\Big|_{\textnormal{All possible contractions}} \to i\gam_{1234} \bar v_1 u_2 \bar v_3 u_4 ~,\label{matrixpsi} 
\end{align}
 with the analytic expression for the vertex,
\begin{align}
i\gam_{1234} \equiv 2 iC^{f_1f_2f_3f_4} \Gamma_{s_1 s_2}\Gamma_{s_3 s_4} - 2i C^{f_1f_4f_3f_2} \Gamma_{s_1 s_4}\Gamma_{s_3 s_2}~, \label{gammapsi}
\end{align}
and the diagrammatic rule,
\begin{align}
\hspace{-4cm}\left(\hspace{1.5cm}\begin{picture}(0,40)
\ArrowLine(0,0)(-20,-20)
\ArrowLine(-20,20)(0,0)
\ArrowLine(0,0)(20,20)
\ArrowLine(20,-20)(0,0)
\Text(-28,-25)[c]{$\psi^{f_1}_{s_1}$}
\Text(-28,25)[c]{$\psi^{f_2}_{s_2}$}
\Text(28,25)[c]{$\psi^{f_3}_{s_3}$}
\Text(28,-25)[c]{$\psi^{f_4}_{s_4}$}
\CCirc(0,0){3}{Black}{Black}
\end{picture}\hspace{1.5cm}\equiv\right) \hspace{2.4cm}\begin{picture}(0,30)
\ArrowLine(0,0)(-20,-20)
\ArrowLine(-20,20)(0,0)
\ArrowLine(0,0)(20,20)
\ArrowLine(20,-20)(0,0)
\Text(-28,-25)[c]{$\psi_1$}
\Text(-28,25)[c]{$\psi_2$}
\Text(28,25)[c]{$\psi_3$}
\Text(28,-25)[c]{$\psi_4$}
\CCirc(0,0){3}{Black}{Black}
\end{picture}\hspace{1.5cm}\longrightarrow \hspace{0.5cm}i\gam_{1234}~,
\label{psivert}      
\end{align}\\[0.cm]
where, as usual, the external states are truncated in the analytic expressions but can be trivially reproduced from the arrows  in the graphs. The same holds for momentum conservation (\ie $\delta$-function) which is suppressed in \eq{matrixpsi} and elsewhere. 

In principle, the same process can be followed for fermion-number violating vertices involving Dirac or Majorana fields (\ie $d=b$). Obviously, after introducing cc-fields to absorb the $\cc$-matrix and using in addition the cc-rearrangement and the Majorana condition when relevant. Some subtle points and technical complications for them will be discussed in Appendix \ref{app:Remarks}. There are first some important remarks in order.

\subsection{Remarks and tools for vertices}
The relative sign between the terms in \eq{gammapsi} is important. It is the well-known Relative Sign of Interfering Feynman (RSIF) graphs which appears here at the vertex. Although the overall sign in a vertex (or an amplitude) is a convention depending on the ordering of ladder operators, the RSIF is not. Neglecting it in fermion diagrams will result in major inconsistencies. Terms will add when they should be substracted and vice versa. Even here one can see clearly the problem. Applying the Fierz identity in \eq{matrixpsi} for $\Gamma\equiv\gamma^\mu P_L$, $f_i=f$ and a relative plus sign in \eq{gammapsi} will make the matrix element vanish identically for any spin and momenta. Obviously, this is wrong.

Another remark concerns the arrow topology  in the graph and the labeling. The clockwise out/in arrow (\ie fermion flow) topology together with the number labels in the graph defines a unique ordering for a fermion \textit{quadrilinear} and a unique correspondence with the number labels in the vertex coupling. To understand how important this ordering is, one needs to go back to first principles.

First recall that a vertex extracted from first order matrix elements as in \eq{gammapsi}, is nothing more than a \textit{fully symmetrized} coupling. It can equivalently replace\footnote{Together with the vertex symmetry factor.} the Lagrangian coupling to any order in the Dyson expansion. The difference is that for the vertex-coupling, contrary to the one in \eq{Lpsi}, all possible contractions for the same diagram are equal. As such, a vertex can also reproduce itself through,
\begin{align}
{1\over 4} ~i\mathbb{\Gamma}_{1'2'3'4'}~ \bra{0} T[ \bar \psi_{1'} \psi_{2'} \bar \psi_{3'} \psi_{4'}] b_4^\dag d_3^\dag b_2^\dag d_1^\dag\ket{0} & = {1\over 4} ~i\mathbb{\Gamma}_{1'2'3'4'}~ \bra{0}  (\bar \psi_{1'} \psi_{2'} \bar \psi_{3'} \psi_{4'}) b_4^\dag d_3^\dag b_2^\dag d_1^\dag\ket{0}\Big|{\begin{array}{c}
\\
\textnormal{\scriptsize{All possible}} \\[-2mm] 
\textnormal{\scriptsize{contractions}}
\end{array} } \nonumber \\
&= + i\mathbb{\Gamma}_{1'2'3'4'}~ \bra{0}  (\wick{\c{\bar \psi_{1'}}\c{d_1^\dag})~  (\c{\psi_{2'}} \c{b_2^\dag})~(\c{\bar \psi_{3'}} \c{d_3^\dag}) ~(\c{\psi_{4'}}\c{b_4^\dag}) }  \ket{0} \nonumber\\
&\to + i\mathbb{\Gamma}_{1234}~ \bar v_1  u_2 \bar v_3 u_4 ~,\nonumber
\end{align}
where $S=4$ is the symmetry factor of the vertex, coming from permutation of identical legs. As for scalars, it is always canceled by the equivalent contractions (\ie four equal contractions here). This is a consistency check for fermion Feynman Rules and eventually produces scalar-type combinatorics for the diagrams. The method above is  valuable. Neglecting symmetry factors and intermediate steps one can trivially: i) understand the reference order set (by someone else) for a given vertex; ii) replace incoming antifermions for outgoing fermions in the vertices and vice versa; iii) set some other reference order for the ladder operators. All these without mistakes and inconsistencies. Keeping track of signs and ordering conventions in four-fermion vertices is crucial, also when performing diagrammatic matching to a UV-theory. 

Notice now the unique correspondence of number labels in couplings and  quadrilinears (expression above) and the clockwise arrow topology and field labeling (vertex-graph). This correspondence is important to be preserved. If not, inconsistencies will arise. Recall that the order for  fermions always matters. Fermion operators (\ie fields and ladder operators) anticommute, irrespective of type and quantum numbers, as a direct consequence of Fermi-statistics. It is only this fundamental property that is responsible for the signs in fermion loops and permutations of legs.

\subsection{Diagrammatic ambiguities and generalization imperatives}
Some of the reasons for a non-trivial generalization of the renormalizable approach may have become clear from the preceding discussion. But there are many more. Nonetheless, all of them are different aspects of two fundamental properties arising in four-fermion (or higher) theories: tensors and Fermi-statistics.

In renormalizable theories fermionic couplings are matrices and as a result the index contraction is non-ambiguous. Therefore, one can suppress  indices and replace them with some orientation convention for reading the diagram. For example, the proper spinorial contraction between propagators and Dirac matrix structures is consistently reproduced by setting a continuous arrow flow and reading the diagram opposite to the arrow direction in Ref.\cite{Denner:1992vza}.

Such compact diagrammatic shortcuts, however, cannot be trivially used here without creating ambiguities. By definition, the four-fermion vertices are tensorial couplings contracted with fermion fields in the Lagrangian. Contrary to matrices, for tensors the multiplication is not unique and must be given explicitly in terms of index contraction. Equivalently, in the diagrams there is more than one possible reading choice as,
$$\begin{picture}(40,60)
\CCirc(0,0){2}{Black}{Black}
\CCirc(40,0){2}{Black}{Black}
\Text(22,50)[c]{(Reading Orientation: \Large ? \normalsize)}
\ArrowLine(0,0)(-20,-20)

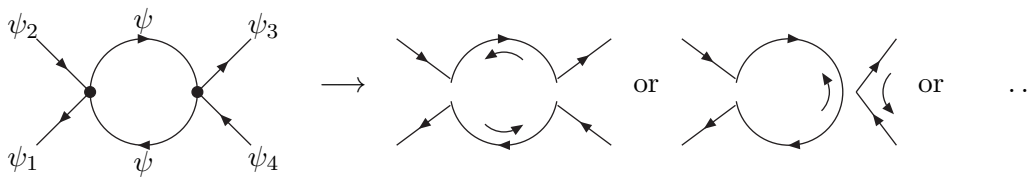
\ArrowLine(-20,20)(0,0)

\ArrowArcn(20,0)(20,0,180)
\ArrowArcn(20,0)(20,180,0)
\ArrowLine(40,0)(60,20)

\ArrowLine(60,-20)(40,0)

\Text(-25,-25)[c]{$\psi_1$}
\Text(-25,25)[c]{$\psi_2$}
\Text(65,25)[c]{$\psi_3$}
\Text(65,-25)[c]{$\psi_4$}
\Text(20,27)[c]{$\psi$}
\Text(20,-27)[c]{$\psi$}
\end{picture}
 \hspace{1.5cm}\longrightarrow \hspace{1.cm}\begin{picture}(40,60)
\LongArrowArc(20,5)(10,45,135)
\LongArrowArc(20,-5)(10,-135,-45)
\ArrowLine(0,-5)(-20,-20)
\ArrowLine(-20,20)(0,5)
\ArrowArcn(20,0)(20,-10,-170)
\ArrowArcn(20,0)(20,170,10)
\ArrowLine(40,5)(60,20)
\ArrowLine(60,-20)(40,-5)
\end{picture} \hspace{1cm}\textnormal{or} \hspace{1.cm}\begin{picture}(40,60)
\ArrowLine(0,-5)(-20,-20)
\ArrowLine(-20,20)(0,5)
\LongArrowArc(25,0)(10,-45,45)
\LongArrowArc(65,0)(10,135,235)
\ArrowArcn(20,0)(20,0,-170)
\ArrowArcn(20,0)(20,170,0)
\ArrowLine(45,0)(60,20)
\ArrowLine(60,-20)(45,0)
\end{picture} \hspace{1cm}\textnormal{or}~~~~~~\dots $$\\[1cm]
\noindent
Nevertheless, all Feynman Rules have been given here with explicit indices for this reason. One can blindly (but carefully in the index order) replace all expressions for propagators, vertices and external states and see how the proper contractions arise. When doing so, it is interesting to notice that all these \textit{``legitimate"} reading paths shown above are consistently reproduced from the spinor structure in the vertices, and nothing more.  But there is another  deeper problem also related to the previous; the  \textit{Fermi-statistics sign}. 

It seems initially striking that the Fermi-sign cannot be included in a graph with the usual diagrammatic rules. More than that, ambiguities  appear in the most unexpected ways. Taking the two-point function for four-fermions by attaching the vertex legs together, different legitimate choices can give two seemingly inequivalent expressions. They read,
\begin{align*}
\begin{picture}(40,40)
\LongArrow(-0.5,19.3)(-1.5,19.3)
\LongArrow(-15,-12)(-23,-12)
\LongArrow(23,-12)(15,-12)
\LongArrow(-14,-17)(-15,-17)
\LongArrow(15,-17)(14,-17)
\CArc(9,3)(20,125,240)
\CArc(10,3)(20,250,270)
\CArc(-10,3)(20,-90,55)
\Line(30,-17)(10.,-17)
\Line(-30,-17)(-10.,-17)
\CCirc(0,-14){3}{Black}{Black}
\Text(0,27)[c]{$\psi$}
\Text(-30,-25)[c]{$\psi_1$}
\Text(30,-25)[c]{$\psi_4$}
\Text(60,-8)[c]{$\longrightarrow$}
\end{picture}\hspace{2cm} &
i\mathcal M^{(1)} \sim i^2\gam_{12'3'4} \bar u_1 u_4 S_{2'3'}\\
& i\mathcal M^{(2)} \sim i^2\gam_{143'2'} \bar u_1 u_4 S_{2'3'}~.
\end{align*}
\\
Assuming for simplicity the one-family case ($f_i =1$) and relabeling indices properly as required for the given matrix element, one naively finds,
\begin{align}
i\mathcal M^{(1)}=2i^2 C^{1111}\int {d^d k\over (2 \pi)^d} \Big( \bar u \Gamma S(k) \Gamma  u  -(\bar u \Gamma u) \Tr [\Gamma S(k)]\Big)= -i\mathcal M^{(2)}~,
\end{align}
and although an RSIF between the terms is present (as should) the overall sign is ambiguous. Taking a fermion-loop sign cannot resolve anything since it will flip signs in both expressions.  

Clearly this is an inconsistency. Under a given set of conventions, any legitimate choice for a diagram (\ie leg relabeling) must give the same result. Moreover, analogous effects can appear internally in diagrams in far more complicated ways and one may never notice the mistake. As usual, the Dyson expansion clears things out. Setting the reference order for creation and annihilation operators and using \eqs{Csym}{gammapsi}, one obtains, 
\begin{align}
{1\over 4} ~i\mathbb{\Gamma}_{1'2'3'4'}~ \bra{0} b_1( \bar \psi_{1'} \psi_{2'} \bar \psi_{3'} \psi_{4'}) b_4^\dag \ket{0} \Big|{\begin{array}{c}
\\
\textnormal{\scriptsize{All possible}} \\[-2mm] 
\textnormal{\scriptsize{contractions}}
\end{array} } & = {1\over 4} i\mathbb{\Gamma}_{1'2'3'4'}~ \bra{0}  (\wick{\c{b_1}\c{\bar \psi_{1'}})~  (\c{\psi_{2'}} \c{\bar \psi_{3'}} ) ~(\c{\psi_{4'}}\c{b_4^\dag}) }  \ket{0} \nonumber \\
&- {1\over 4} i\mathbb{\Gamma}_{1'2'3'4'}~ \bra{0}  (\wick{\c{b_1}\c{\bar \psi_{1'}})~ (\c{\psi_{2'}}\c{b_4^\dag}) ~(\c{\psi_{4'}}\c{\bar \psi_{3'}}  )  }  \ket{0} \nonumber\\
 &+ {1\over 4} ~~\dots ~~ - {1\over 4} ~~ \dots \nonumber\\
 \to & + i^2\gam_{12'3'4} \bar u_1 u_4 S_{2'3'}\nonumber = - i^2\gam_{143'2'} \bar u_1 u_4 S_{2'3'} ~,\nonumber 
\end{align}
so that the ambiguity is resolved and only $i\mathcal{M}^{(1)}$ is correct. The fixed order of the quadrilinear and the reference order of the ladder operators uniquely determined the sign.

\section{Amplitude Calculations}\label{sec:amplitude}

For practical calculations the Wick-contraction approach cannot obviously withstand. Thus, a very careful algorithm needs to be devised. Although the algorithm here relies also on a Wick-step, the main point is that it applies once for whole classes of diagrams, neglecting symmetry factors. This minor inconvenience is eventually its main strength. It results into a compact description with no fundamental distinction from the scalar case. The advantages of this method will become more clear when discussing the standard alternatives, in the next section.

The following algorithm applies for amplitude calculations involving four-fermion vertices. For diagrams without four-fermions the method is equivalent to the purely diagrammatic prescriptions of Ref.\cite{Denner:1992vza,Denner:1992me,Peskin:1995ev}. One can always switch between the two if more convenient.  

\subsection{An algorithmic approach}\label{sec:algorithm}
Assuming that the theoretical framework in the EFT-Lagrangian adopts the formalism of Ref.\cite{Denner:1992vza} and the vertex setup follows the guidelines of \eqs{Lpsi}{matrixpsi} setting a unique correspondence with the graph topology of \eqref{psivert} (as in Ref.\cite{Dedes:2017zog}). Recalling also that summation over primed indices  is always considered (Fig.\ref{fig:not}), then: 
\begin{enumerate}
\item Set a reference order for creation/annihilation operators which must be kept unchanged for all diagrams to a given process. Assign them number labels and use the same labels on external legs. For Majorana and  Dirac fermion-number violating apply fermion arrows at the next step.
\begin{align*}
\begin{picture}(80,80)
\Text(78,60)[c]{Example: $\psi_2 \psi_4 \to \psi_1\psi_3 $ with vertex of Section \ref{sec:vert} to one-loop order.}
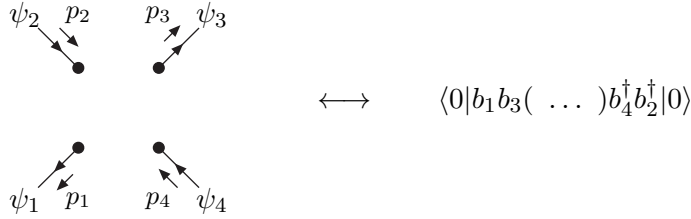
\ArrowLine(-15,-15)(-30,-30)
\Line(-17,-25)(-22,-30)
\ArrowLine(-21,-29)(-22,-30)
\ArrowLine(-30,30)(-15,15)
\Line(-17,25)(-22,30)
\ArrowLine(-17,25)(-16,24)
\ArrowLine(15,15)(30,30)
\Line(17,25)(22,30)
\ArrowLine(21,29)(22,30)
\ArrowLine(30,-30)(15,-15)
\Line(17,-25)(22,-30)
\ArrowLine(17,-25)(16,-24)
\Text(-35,-35)[c]{$\psi_1$}
\Text(-15,-35)[c]{\small{$p_1$}}
\Text(-35,35)[c]{$\psi_2$}
\Text(-15,35)[c]{\small{$p_2$}}
\Text(35,35)[c]{$\psi_3$}
\Text(15,35)[c]{\small{$p_3$}}
\Text(35,-35)[c]{$\psi_4$}
\Text(15,-35)[c]{\small{$p_4$}}
\CCirc(15,15){2}{Black}{Black}
\CCirc(15,-15){2}{Black}{Black}
\CCirc(-15,15){2}{Black}{Black}
\CCirc(-15,-15){2}{Black}{Black}
\end{picture}\hspace{-0.3cm}\longleftrightarrow ~~~~~~\bra{0} b_1 b_3 ( ~\dots ~ ) b_4^\dag b_2^\dag \ket{0}
~~~~~~~\\[1.5cm]
\end{align*}

\item Write down all topologically distinct diagrams contributing to a given process by attaching the vertices together. While doing so, \textit{keep the arrow topologies of vertices intact, \eg do not twist legs}. Assign number labels for all internal vertex legs. 
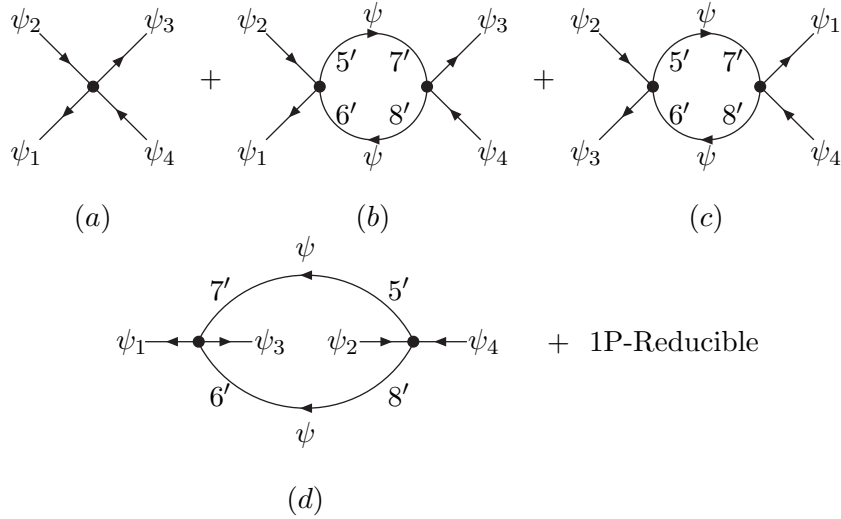
\begin{align*}
 ~~~~~~~~~~\begin{picture}(80,40)
\Text(0,-50)[c]{$(a)$}
\CCirc(0,0){2}{Black}{Black}
\ArrowLine(0,0)(-20,-20)
\ArrowLine(-20,20)(0,0)
\ArrowLine(0,0)(20,20)
\ArrowLine(20,-20)(0,0)
\Text(-25,-25)[c]{$\psi_1$}
\Text(-25,25)[c]{$\psi_2$}
\Text(25,25)[c]{$\psi_3$}
\Text(25,-25)[c]{$\psi_4$}
\end{picture}
\hspace{-1.5cm}+\hspace{1.2cm}
\begin{picture}(80,0)
\Text(20,-50)[c]{$(b)$}
\CCirc(0,0){2}{Black}{Black}
\CCirc(40,0){2}{Black}{Black}
\ArrowLine(0,0)(-20,-20)
\ArrowLine(-20,20)(0,0)
\ArrowArcn(20,0)(20,0,180)
\ArrowArcn(20,0)(20,180,0)
\ArrowLine(40,0)(60,20)
\ArrowLine(60,-20)(40,0)
\Text(-25,-25)[c]{$\psi_1$}
\Text(-25,25)[c]{$\psi_2$}
\Text(65,25)[c]{$\psi_3$}
\Text(65,-25)[c]{$\psi_4$}
\Text(10,10)[c]{$5'$}
\Text(10,-10)[c]{$6'$}
\Text(30,10)[c]{$7'$}
\Text(30,-10)[c]{$8'$}
\Text(20,27)[c]{$\psi$}
\Text(20,-27)[c]{$\psi$}
\end{picture}
\hspace{-0.1cm}+\hspace{1.2cm}
\begin{picture}(80,0)
\Text(20,-50)[c]{$(c)$}
\CCirc(0,0){2}{Black}{Black}
\CCirc(40,0){2}{Black}{Black}
\ArrowLine(0,0)(-20,-20)
\ArrowLine(-20,20)(0,0)
\ArrowArcn(20,0)(20,0,180)
\ArrowArcn(20,0)(20,180,0)
\ArrowLine(40,0)(60,20)
\ArrowLine(60,-20)(40,0)
\Text(-25,-25)[c]{$\psi_3$}
\Text(-25,25)[c]{$\psi_2$}
\Text(65,25)[c]{$\psi_1$}
\Text(65,-25)[c]{$\psi_4$}
\Text(10,10)[c]{$5'$}
\Text(10,-10)[c]{$6'$}
\Text(30,10)[c]{$7'$}
\Text(30,-10)[c]{$8'$}
\Text(20,27)[c]{$\psi$}
\Text(20,-27)[c]{$\psi$}
\end{picture}
\\[2.8cm] 
\begin{picture}(250,0)
\Text(40,-60)[c]{$(d)$}
\CCirc(0,0){2}{Black}{Black}
\CCirc(80,0){2}{Black}{Black}
\ArrowLine(0,0)(-20,0)
\ArrowLine(0,0)(20,0)
\ArrowLine(60,0)(80,0)
\ArrowLine(100,0)(80,0)
\ArrowArcn(40,20)(45,-25,-155)
\ArrowArc(40,-20)(45,25,155)
\Text(-25,0)[c]{$\psi_1$}
\Text(25,0)[c]{$~\psi_3$}
\Text(55,0)[c]{$\psi_2\,$}
\Text(105,0)[c]{$~\psi_4$}
\Text(10,20)[c]{$7'~$}
\Text(75,20)[c]{$5'$}
\Text(10,-20)[c]{$6'~$}
\Text(75,-20)[c]{$8'$}
\Text(40,35)[c]{$\psi$}
\Text(40,-35)[c]{$\psi$}
\Text(170,0)[c]{$+~$ 1P-Reducible}
\end{picture}\\[2cm]
\end{align*}
\item Clashing arrows should never appear. When fermion-number violating vertices are relevant, clashing arrows can be avoided by considering the appropriate cc-vertices, instead. They can be derived from the Lagrangian through the bilinear property (see Section \ref{sec:fflow}),
\begin{align*}
\bar\chi_{1'} \mathbb{\Gamma}_{1'2'} \xi^c_{2'} &= \bar\xi_{1'} \mathbb{\Gamma}^c_{1'2'} \chi^c_{2'}~.
\end{align*}
\item Group the diagrams into classes: diagrams which are identical under the interchange of final (or initial) states belong to the same class.
\begin{align*}
\textnormal{Class-1} &= \{ (a)\}~,\\
\textnormal{Class-2} &= \{(b),(c)\}~,\\
\textnormal{Class-3} &=  \{(d)\}~.
\end{align*}
\item Choose one diagram from each class and write the analytic expression replacing blindly through \eqst{S}{psiextc} and \eqref{psivert}. Be careful with order of indices. Due to permutation/relabeling of identical legs there can be many legitimate choices, and any will do. 
\begin{align*}
\begin{picture}(80,40)
\Text(-63,0)[c]{$(a)$}
\CCirc(0,0){2}{Black}{Black}
\LongArrow(-7,-15)(-12,-20)
\LongArrow(-12,20)(-7,15)
\LongArrow(7,15)(12,20)
\LongArrow(12,-20)(7,-15)
\ArrowLine(0,0)(-20,-20)
\ArrowLine(-20,20)(0,0)
\ArrowLine(0,0)(20,20)
\ArrowLine(20,-20)(0,0)
\Text(-25,-25)[c]{$\psi_1$}
\Text(-25,25)[c]{$\psi_2$}
\Text(25,25)[c]{$\psi_3$}
\Text(25,-25)[c]{$\psi_4$}
\end{picture}&\longrightarrow ~ i\mathcal{M}_{(a)}\sim i \mathbb{\Gamma}_{1234}~\bar u_1 u_2 \bar u_3 u_4~,\\[1.5cm]
\begin{picture}(100,40)
\Text(-44,0)[c]{$(b)$}
\CCirc(0,0){2}{Black}{Black}
\CCirc(40,0){2}{Black}{Black}
\ArrowLine(0,0)(-20,-20)
\Line(-7,-15)(-12,-20)
\ArrowLine(-11,-19)(-12,-20)
\ArrowLine(-20,20)(0,0)
\Line(-7,15)(-12,20)
\ArrowLine(-7,15)(-6,14)
\ArrowArcn(20,0)(20,0,180)
\ArrowArcn(20,0)(20,180,0)
\ArrowLine(40,0)(60,20)
\Line(47,15)(52,20)
\ArrowLine(51,19)(52,20)
\ArrowLine(60,-20)(40,0)
\Line(47,-15)(52,-20)
\ArrowLine(47,-15)(46,-14)
\Text(-25,-25)[c]{$\psi_1$}
\Text(-25,25)[c]{$\psi_2$}
\Text(65,25)[c]{$\psi_3$}
\Text(65,-25)[c]{$\psi_4$}
\Text(10,10)[c]{$5'$}
\Text(10,-10)[c]{$6'$}
\Text(30,10)[c]{$7'$}
\Text(30,-10)[c]{$8'$}
\Text(20,27)[c]{$\psi$}
\Text(20,-27)[c]{$\psi$}
\end{picture}
&\longrightarrow ~i\mathcal{M}_{(b)}\sim i^4 \mathbb{\Gamma}_{125'6'} \mathbb{\Gamma}_{8'7'34} ~S_{7'5'} S_{6'8'} ~\bar u_1 u_2 \bar u_3 u_4~,\\[1.5cm]
\begin{picture}(120,40)
\Text(-25,-25)[c]{$(d)$}
\CCirc(0,0){2}{Black}{Black}
\CCirc(80,0){2}{Black}{Black}
\LongArrow(10,5)(18,5)
\LongArrow(-10,5)(-18,5)
\LongArrow(98,5)(90,5)
\LongArrow(62,5)(70,5)
\ArrowLine(0,0)(-20,0)
\ArrowLine(0,0)(20,0)
\ArrowLine(60,0)(80,0)
\ArrowLine(100,0)(80,0)
\ArrowArcn(40,20)(45,-25,-155)
\ArrowArc(40,-20)(45,25,155)
\Text(-25,0)[c]{$\psi_1$}
\Text(25,0)[c]{$~\psi_3$}
\Text(55,0)[c]{$\psi_2\,$}
\Text(105,0)[c]{$~\psi_4$}
\Text(10,20)[c]{$7'~$}
\Text(75,20)[c]{$5'$}
\Text(10,-20)[c]{$6'~$}
\Text(75,-20)[c]{$8'$}
\Text(40,35)[c]{$\psi$}
\Text(40,-35)[c]{$\psi$}
\end{picture}& \longrightarrow ~i\mathcal{M}_{(d)}\sim i^4 \mathbb{\Gamma}_{17'36'} \mathbb{\Gamma}_{8'25'4} ~S_{7'5'} S_{6'8'} ~\bar u_1 \bar u_3 u_2  u_4~.\\[2cm]
\end{align*}
The order of the (commuting) external Dirac-spinors is always  irrelevant. 

\item Do not add signs due to Fermi-statistics \ie fermion loops and/or  leg interchange. Instead, reproduce each expression of step-5 through a single Wick contraction. This will determine the \textit{overall} sign of the diagram. For the contractions, always start from the ordering convention of step-1 and notice that \textit{the ordering of the fermion quadrilinear is predetermined by the vertex graph}. 
\begin{align*}
i\mathcal{M}_{(a)}&\sim + i \mathbb{\Gamma}_{1'2'3'4'}
\bra{0}
\wick{ (\c{b_1}\c{\bar \psi_{1'}})(\c{\psi_{2'}}\c{b_2^\dag})  
(\c{b_3}\c{\bar \psi_{3'}}) (\c{\psi_{4'}} \c{b_4^\dag})}
 \ket{0} \nonumber\\
&\sim +  i \mathbb{\Gamma}_{1234} ~\bar u_1 u_2 \bar u_3 u_4~,
\\
i\mathcal{M}_{(b)}&\sim + i^2 \mathbb{\Gamma}_{1'2'5'6'}\mathbb{\Gamma}_{8'7'3'4'}
\bra{0}
[\,\wick{(\c{b_1}\c{\bar \psi_{1'}})(\c{\psi_{2'}}\c{b_2^\dag})  
~\c2{\bar \psi_{5'}} \c{\psi_{6'}} \,]~[\, \c{\bar\psi_{8'}} \c2{\psi_{7'}}~ (\c{b_3}\c{\bar \psi_{3'}})(\c{\psi_{4'}} \c{b_4^\dag})}\,]
 \ket{0}
\nonumber\\
&\sim - i^4 \mathbb{\Gamma}_{125'6'} \mathbb{\Gamma}_{8'7'34} ~S_{7'5'} S_{6'8'} ~\bar u_1 u_2 \bar u_3 u_4 ~,\nonumber
\\
i\mathcal{M}_{(d)}&\sim - i^2 \mathbb{\Gamma}_{1'7'3'6'}\mathbb{\Gamma}_{8'2'5'4}
\bra{0}
[\,\wick{(\c{b_1}\c{\bar \psi_{1'}})\c2{\psi_{7'}}(\c{b_3}\c{\bar \psi_{3'}})\c1{\psi_{6'}} ~]~[~ \c{\bar\psi_{8'}}(\c{\psi_{2'}}\c{b_2^\dag})~\c2{\bar \psi_{5'}}~(\c{\psi_{4'}} \c{b_4^\dag})}\,]
 \ket{0}
\nonumber\\
&\sim - i^4 \mathbb{\Gamma}_{17'36'} \mathbb{\Gamma}_{8'25'4} ~S_{7'5'} S_{6'8'} ~\bar u_1 \bar u_3 u_2  u_4 ~.\nonumber
\end{align*}

\item Determine the symmetry factor diagrammatically as for scalars: Majorana fermions behave as real scalars, Dirac fermions as charged scalars and when cc-fields are relevant, $\psi^c\sim \bar\psi$ and $\bar\psi^c\sim \psi$, for counting\footnote{Note that the same combinatorics apply also for two-fermion vertices in the standard approach. For example, recall the usual $(-{1\over 2})$ factor for a Majorana-neutrino loop.}. Topologically distinct graphs are also understood in the same ``scalar" sense.
\bea
S_{(a)}= S_{(b)} =1 ~,~~ S_{(d)}=2 ~.  \nonumber
\eea
\item Apply momentum conservation and use the analytic expressions for vertices (as in \eq{gammapsi}). Momentum and spin-labels for external spinors  can be applied at the end.
\begin{align*}
i\mathcal{M}_{(a)}& = + ~\delta(p_1+p_3-p_2-p_4) ~ i \mathbb{\Gamma}_{1234} ~\bar u_1 u_2 \bar u_3 u_4~,\\
i\mathcal{M}_{(b)}& = - ~ \delta(p_1+p_3-p_2-p_4) ~\Big[\bar u_1 u_2  \bar u_3  u_4 ~ \times \\
 &~~~\int{d^dk\over (2\pi)^d} i^4 \mathbb{\Gamma}_{125'6'} \mathbb{\Gamma}_{8'7'34} ~S_{7'5'} (k+p_2-p_1) S_{6'8'}(k)\Big] \\
&=\dots\dots~,
\\
i\mathcal{M}_{(d)}& = -{1\over 2} ~ \delta(p_1+p_3-p_2-p_4) ~\Big[\bar u_1 u_2  \bar u_3  u_4 ~ \times \\
 &~~~\int{d^dk\over (2\pi)^d} i^4 \mathbb{\Gamma}_{17'36'} \mathbb{\Gamma}_{8'25'4} ~S_{7'5'} (k+p_2+p_4) S_{6'8'}(-k)\Big] \\
&=\dots\dots 
\end{align*}
\item Once a diagram in a class is determined, the other topologically distinct graphs in the same class can be produced instead with the usual shortcut\footnote{See for example Peskin\&Schroeder\cite{Peskin:1995ev}, p.120 and Ref.\cite{Denner:1992vza} for more complicated cases.}. 

Permuting the relevant external indices in the expression for $i\mathcal{M}_{(b)}$ above and considering \textit{now} an additional overall minus sign (\ie RSIF), one trivially obtains,
\begin{align*}
~~~~~~~~~\begin{picture}(80,40)
\Text(20,-50)[c]{$(c)$}
\CCirc(0,0){2}{Black}{Black}
\CCirc(40,0){2}{Black}{Black}
\ArrowLine(0,0)(-20,-20)
\Line(-7,-15)(-12,-20)
\ArrowLine(-11,-19)(-12,-20)
\ArrowLine(-20,20)(0,0)
\Line(-7,15)(-12,20)
\ArrowLine(-7,15)(-6,14)
\ArrowArcn(20,0)(20,0,180)
\ArrowArcn(20,0)(20,180,0)
\ArrowLine(40,0)(60,20)
\Line(47,15)(52,20)
\ArrowLine(51,19)(52,20)
\ArrowLine(60,-20)(40,0)
\Line(47,-15)(52,-20)
\ArrowLine(47,-15)(46,-14)
\Text(-25,-25)[c]{$\psi_3$}
\Text(-25,25)[c]{$\psi_2$}
\Text(65,25)[c]{$\psi_1$}
\Text(65,-25)[c]{$\psi_4$}
\Text(10,10)[c]{$5'$}
\Text(10,-10)[c]{$6'$}
\Text(30,10)[c]{$7'$}
\Text(30,-10)[c]{$8'$}
\Text(20,27)[c]{$\psi$}
\Text(20,-27)[c]{$\psi$}
\end{picture}
\hspace{-0.8cm}
\begin{array}{rl}
 & \\
 & \\
\longrightarrow 
i\mathcal{M}_{(c)} = & (+)  ~ \delta(p_3+p_1-p_2-p_4) ~\Big[\bar u_3 u_2  \bar u_1  u_4 ~ \times \\
 &  \int{d^dk\over (2\pi)^d} i^4 \mathbb{\Gamma}_{325'6'} \mathbb{\Gamma}_{8'7'14} ~S_{7'5'} (k+p_2-p_3) S_{6'8'}(k)\Big]  
\end{array}\\[0.7cm]
\end{align*}

\item One should refer to Appendix \ref{app:Remarks} for diagrams involving the four-Majorana vertex. Some special remarks for fermion-number violating vertices are also given there.
\end{enumerate}
\noindent

It is important not to deviate from the algorithm and take all relevant steps even in the most trivial diagrams. The anticommuting quadrilinears, always underlying the four-fermion graphs, have a tricky nature. Performing diagrammatic shortcuts without being aware of their implications on the Wick-contractions will easily lead to inconsistencies, as the ones discussed in previous sections and many more. 

\subsection{Comparison with the traditional EFT diagrams}\label{sec:comparison}
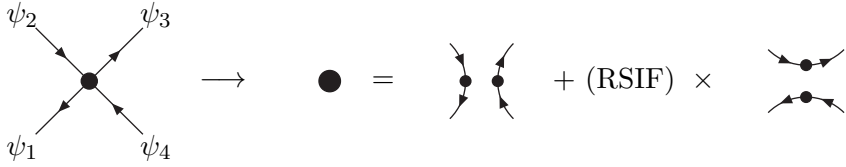
\begin{figure}[t]
\begin{center}
\hspace{2cm}\begin{picture}(40,20)
\ArrowLine(0,0)(-20,-20)
\ArrowLine(-20,20)(0,0)
\ArrowLine(0,0)(20,20)
\ArrowLine(20,-20)(0,0)
\Text(-25,-25)[c]{$\psi_1$}
\Text(-25,25)[c]{$\psi_2$}
\Text(25,25)[c]{$\psi_3$}
\Text(25,-25)[c]{$\psi_4$}
\CCirc(0,0){3}{Black}{Black}
\Text(50,0)[c]{$\longrightarrow$}
\end{picture}
\hspace{1.5cm} 
\begin{picture}(80,0)
\CCirc(0,0){4}{Black}{Black}
\Text(20,0)[c]{$=$}
\end{picture}
\hspace{-2.2cm} \hspace{1cm}
\begin{picture}(80,0)
\ArrowArcn(-26,0)(20,0,-45)
\ArrowArcn(-26,0)(20,45, 0)
\ArrowArcn(26,0)(20,225,180)
\ArrowArcn(26,0)(20,180,135)
\CCirc(-6,0){2}{Black}{Black}
\CCirc(6,0){2}{Black}{Black}
\Text(60,0)[c]{$+~ \textnormal{(RSIF)} ~\times~$} 
\end{picture}
\hspace{1.2cm}
\begin{picture}(80,0)
\ArrowArc(0,26)(20,225,270)
\ArrowArc(0,26)(20,270,315)
\ArrowArc(0,-26)(20,90,135)
\ArrowArc(0,-26)(20,45,90)
\CCirc(0,6){2}{Black}{Black}
\CCirc(0,-6){2}{Black}{Black}
\end{picture}\vspace{1cm}
\caption{Typical decomposition of the vertices into fermion-flow pairs. The latter are the traditional EFT diagrams which are dropped in the present formalism.}
\label{fig:arrowpair}
\end{center}
\end{figure}
The typical diagrammatic picture one obtains from the four-fermion vertices introduced here (\ie matrix elements) is given in Fig.\ref{fig:arrowpair},
with relative signs due to anticommutation (RSIF) already incorporated.
As should be clear, only the LHS vertex is required for amplitude calculations. The RHS, which also reflects the spinorial structures in \eq{gammapsi}, shows how the tensorial vertices decompose into diagrams of fermion-flow pairs. It is also the commonly used diagrammatic language of EFT.

In Weak Effective Theory (WET), in particular, these vertices have been extensively used. Such objects deviate from the standard definition of a vertex, at least in the usual renormalizable sense. They are partially symmetrized couplings\footnote{Even worse, one may define the WET-vertex as the completely unsymmetrized Lagrangian coupling.} and thus cannot reproduce themselves \textit{trivially} in the Dyson expansion. They are consistent but they are not much of a shortcut as compared to the initial Lagrangian coupling. 

The WET-vertex related to the first term in \eq{gammapsi} and reflecting the first arrow-pair topology above, reads,
\begin{align*}
i\gam^{W}_{1234} = 2i C^{f_1f_2f_3f_4} \Gamma _{s_1s_2}\Gamma _{s_3s_4}~.
\end{align*}
As any consistent coupling for the Lagrangian, symmetrized or not, it can participate in the Dyson expansion with the suitable symmetry factor. Here,  
\begin{align*}
{1\over 2} ~i\gam^{W}_{1'2'3'4'}~ \bra{0} ( \bar \psi_{1'} \psi_{2'} \bar \psi_{3'} \psi_{4'})b_4^\dag d_3^\dag b_2^\dag d_1^\dag \ket{0} \Big|{\begin{array}{c}
\\
\textnormal{\scriptsize{All possible}} \\[-2mm] 
\textnormal{\scriptsize{contractions}}
\end{array} } & \to ~(i\gam^{W}_{1234} - i\gam^{W}_{1432}) \bar v_1 u_2 \bar v_3 u_4 \nonumber \\
& =~  i\gam_{1234}  \bar v_1 u_2 \bar v_3 u_4 ~,
\end{align*}
producing the second arrow topology above and eventually the matrix element (\ie vertex) of \eq{matrixpsi}, with considerable effort.

It is not hard to imagine that the arrow-pair approach can easily lead to mistakes concerning signs, factors and missing terms/diagrams in amplitudes. Conversely, the algorithm presented previously has no such difficulties. Symmetry factors are read as for scalars, relative signs are already implemented through the vertices and the overall sign is obtained from a single Wick contraction for whole classes of diagrams. More importantly, for each diagram with n-number of four-fermion vertices one needs a single diagram here versus $2^n$ diagrams (and thus $2^n$  possibilities for a mistake) in the traditional treatment. 

\section{Conclusions}\label{sec:conclusions}
A compact method for deriving amplitudes involving four-fermion or higher operators has been presented. The main points of the approach can be summarized as follows:
\begin{itemize}
\item The basic formalism of Denner \textit{et al.,} \cite{Denner:1992vza,Denner:1992me} is sufficient to gereralize the matrix treatment of renormalizable theories into a tensor treatment for four-fermions or higher (see Section \ref{sec:basic}).
\item Some modifications in the diagrammatic treatment are necessary due to new aspects one faces with four-fermions. They arise from fundamental mathematical/physical properties \ie \textit{tensor multiplication is not unique, fermion quadrilinears anticommute and appear in the first order of the Dyson expansion} (see Section \ref{sec:generalization}).
\item  The method relies on a suitable vertex setup which is a trivial generalization of the one used in renormalizable theories. Non-trivial modifications appear in the reading rules of the diagram: dropping reading orientation, keeping the arrow topology of vertices intact, applying scalar-type treatment and adding a Wick-step once for whole classes of diagrams, to determine the \textit{overall} sign due to Fermi-statistics (see Section \ref{sec:algorithm}). 
\item As compared to the traditional method of fermion-flow pairs (\ie unattached vertices of Fig. \ref{fig:arrowpair}) the method possesses advantages with respect to signs, factors and possible omission of diagrams. Each diagram here typically corresponds to $2^n$ diagrams for the alternative, with $n$ being the number of four-fermion vertices in a given diagram (see Section \ref{sec:comparison}).
\end{itemize} 
Finally a comment, also suitable for code implementations. The Wick-contraction picture employed here repeatedly is the algebraic and fundamental way to perform amplitude calculations. As such, implementing a quadrilinear of the form\footnote{Dots here reflect the fact that fermion fields form an anticommuting quadrilinear and a minus sign is generated for each permutation. As usual, primed numbers denote summation over sets of relevant indices and $S$ is the symmetry factor of the vertex, depending on the common fermion fields involved (counting follows the combinatorics explained in the algorithm).},
\begin{align*}
{1\over S} i \gam_{1'2'3'4'} (\bar\psi_{1'} . \psi_{2'} .\bar\psi_{3'} . \psi_{4'})~,
\end{align*} 
in code chains treating properly tensors and fermion anticommutation should suffice. Recalling in addition that the diagrams here are treated as scalars when it comes to topologies and combinatorics. The only thing missing is an overall sign, to be determined through anticommutation of fermion operators in order to bring the contractions together and in the order of \eqst{S}{psiextc}. It should be clear also that using the tensor vertices $i\gam$ without implementing the RSIF in their definition can result into major inconsistencies for the final amplitudes.
\section*{Acknowledgments}
The author wishes to thank I.~Brivio and M.~Trott for their kind hospitality during his visit at Niels Bohr I., and for useful discussions and guidance in available computer-codes for amplitude calculations. He would also like to thank A.~Dedes, J.~Rosiek and K.~Suxho for valuable suggestions, remarks and proofreading at various stages of the manuscript. Finally, he wishes to express his gratitude to all members of the Particle Physics Division in Athens U. (EKPA), for their kind hospitality and further to V.~Spanos for intriguing conversations on various aspects, directly or indirectly related to this work. 
\newpage  
\appendix
\section{Remarks for fermion-number violation and four-Majorana}
\label{app:Remarks}
There can be technical difficulties and subtle points when it comes to fully symmetrizing a vertex for Fermion-Number (FN) violating  operators. The problem arises from the fact that the cc-transformation used to absorb/eliminate the $\cc$-dependence is a bilinear operation. On the other hand, the four-fermion vertex is a fully symmetrized \textit{tensor} coupling, obtained by taking all possible contractions with an ordered set of ladder operators on a fermion \textit{quadrilinear}. Therefore, there can be remnants. Conversely, these problems are absent when the cc-operation is irrelevant, as for example in FN-conserving four-fermions or higher. 

Such complications refer to very few vertices of EFTs based on SM, being typically uninterresting due to severe experimental bounds. For example, this is the case in dimension-six SMEFT\cite{Grzadkowski:2010es,Dedes:2017zog}  where they are confronted in the B-violating and the single four-neutrino (\ie Majorana) vertex.

\subsection{The B-violating vertices of SMEFT}
Applying the cc-operation on the Lagrangian, one arrives at the two terms for baryon-number violating operators having the structure,
\begin{align}
\mathcal{L}_{B} \sim &\sum C^{f_1f_2f_3f_4} (P_L)_{s_1s_2} (P_L)_{s_3s_4}
 ~~(\bar u^c)^{f_1}_{s_1} d^{f_2}_{s_2} (\bar u^c)^{f_3}_{s_3} e^{f_4}_{s_4} \nonumber \\
+& \sum C^{f_1f_2f_3f_4} (P_L)_{s_1s_2} (P_L)_{s_3s_4}  
~~(\bar u^c)^{f_1}_{s_1} u^{f_2}_{s_2} (\bar d^c)^{f_3}_{s_3} e^{f_4}_{s_4}~, 
\end{align}
where color is ignored and relevant terms are displayed with a simplified form, as mentioned in Section \ref{sec:introduction}. Treating them as independent vertices, one obtains the Feynman Rules, 
$$
\hspace{-2cm}\begin{picture}(150,40)
\ArrowLine(0,0)(-20,-20)
\ArrowLine(-20,20)(0,0)
\ArrowLine(0,0)(20,20)
\ArrowLine(20,-20)(0,0)
\CCirc(0,0){2}{Black}{Black}
\Text(-25,-25)[c]{$u^c_1$}
\Text(-25,25)[c]{$d_2$}
\Text(25,25)[c]{$u^c_3$}
\Text(25,-25)[c]{$e_4$}
\Text(80,0)[c]{$\longrightarrow i\gam^{(1)}_{1234}~, $}
\end{picture} 
\hspace{1cm}
\begin{picture}(0,40)
\ArrowLine(0,0)(-20,-20)
\ArrowLine(-20,20)(0,0)
\ArrowLine(0,0)(20,20)
\ArrowLine(20,-20)(0,0)
\CCirc(0,0){2}{Black}{Black}
\Text(-25,-25)[c]{$u^c_1$}
\Text(-25,25)[c]{$u_2$}
\Text(25,25)[c]{$d^c_3$}
\Text(25,-25)[c]{$e_4$}
\Text(80,0)[c]{$\longrightarrow i\gam^{(2)}_{1234} ~, $}
\end{picture} \\[1.5cm]
$$
with the simplified vertices\footnote{The full expression including color is given in Arxiv v4., of Ref.\cite{Dedes:2017zog}.} reading explicitly (momenta always incoming),
\begin{align}
i\gam^{(1)}_{1234} = iC^{f_1f_2f_3f_4} (P_L)_{s_1s_2} (P_L)_{s_3s_4}-iC^{f_3f_2f_1f_4} (P_L)_{s_1s_4} (P_L)_{s_3s_2}~,  \\
i\gam^{(2)}_{1234} = iC^{f_1f_2f_3f_4} (P_L)_{s_1s_2} (P_L)_{s_3s_4}+iC^{f_2f_1f_3f_4} (P_L)_{s_1s_2} (P_L)_{s_3s_4} ~,
\end{align} 
corresponding to the ``matrix elements",
\begin{align}
i\gam^{(1)}_{1234} \bar v_{u_1} u_{d_2} \bar v_{u_3} u_{e_4}~ ,~~
i\gam^{(2)}_{1234}  \bar v_{u_1} u_{u_2} \bar v_{d_3} u_{e_4}~, 
\end{align} 
and more generally to the fully symmetrized quadrilinears,
\begin{align}
{1\over 2} i\gam^{(1)}_{1'2'3'4'}  \bar u^c_{1'} d_{2'} \bar u^c_{3'} e_{4'}~ ,~~
{1\over 2} i\gam^{(2)}_{1'2'3'4'}  \bar u^c_{1'} u_{2'} \bar d^c_{3'} e_{4'} ~,
\end{align} 
participating to any order in the Dyson expansion.

These are not different matrix elements in the usual sense since they are associated with spinor/field redefinitions. Applying \eq{ccspinors} or \eqref{psicc} one could in principle merge the terms together but only at the cost of explicit $\cc$-dependence. Equivalently, there is no cc-operation which can merge the vertex graphs together without clashing arrows\footnote{This is better understood from the arrow-pair decomposition of Fig. \ref{fig:arrowpair}.}.

Even so, the vertices here originate from different terms in the Lagrangian. As such, they can be treated as independent with the provision that they always contribute to the same processes (\ie tree or loop) and an RSIF can arise between them. Following the algorithm without deviations (\ie recall the first step), the latter is always taken into account. 

Finally, keep in mind that  cc-fields and cc-vertices are equivalent ways to reexpress the Lagrangian terms. They have a suitable form that avoids clashing arrows in the graphs and, equivalently, the explicit presence of $\cc$ in the amplitudes. Diagrams associated with the cc-operation should be taken only once, to avoid double counting. 

\subsection{The exceptional four-neutrino vertex of SMEFT}
The situation for the four-Majorana vertex is more complex although it arises for the same technical reason. The previous treatment of including two vertices for the same matrix element instead of a merged one with $\cc$ explicit, fails. Here, a single term in the mass basis Lagrangian creates three arrow-pair topologies when symmetrized and one of them cannot be merged without clashing arrows.

Starting as usual from the simplified term in the Lagrangian,
\begin{align}
\mathcal{L}_{4\nu} \sim &\sum C^{f_1f_2f_3f_4} (\gamma_\mu P_L)_{s_1s_2} (\gamma^\mu P_L)_{s_3s_4}
~~ \bar \nu^{f_1}_{s_1} \nu^{f_2}_{s_2} \bar \nu^{f_3}_{s_3} \nu^{f_4}_{s_4} ~,
\end{align}  
one obtains for the vertex (incoming momenta), \\
$$
\begin{picture}(40,30)
\Line(0,0)(-20,-20)
\Line(-20,20)(0,0)
\Line(0,0)(20,20)
\Line(20,-20)(0,0)
\Text(-25,-25)[c]{$\nu_1$}
\Text(-25,25)[c]{$\nu_2$}
\Text(25,25)[c]{$\nu_3$}
\Text(25,-25)[c]{$\nu_4$}
\Text(60,0)[c]{$\longrightarrow i\gam_{1234}$}
\CCirc(0,0){3}{Black}{Black}
\end{picture}\vspace{1.cm}
$$
$$ 
\hspace{1cm}
\begin{picture}(0,40)
\CCirc(0,0){4}{Black}{Black}
\Text(20,0)[c]{$= $}
\end{picture}
\hspace{2cm}
\begin{picture}(80,0)
\ArrowArcn(-26,0)(20,0,-45)
\ArrowArcn(-26,0)(20,45, 0)
\ArrowArcn(26,0)(20,225,180)
\ArrowArcn(26,0)(20,180,135)
\CCirc(-6,0){2}{Black}{Black}
\CCirc(6,0){2}{Black}{Black}
\Text(60,0)[c]{$+~ \textnormal{(RSIF)} ~\times~$} 
\Text(0,-25)[c]{$(1)$}
\Text(83,-25)[c]{$(2)$}
\end{picture}
\hspace{1.2cm}
\begin{picture}(80,0)
\ArrowArc(0,26)(20,225,270)
\ArrowArc(0,26)(20,270,315)
\ArrowArc(0,-26)(20,90,135)
\ArrowArc(0,-26)(20,45,90)
\CCirc(0,6){2}{Black}{Black}
\CCirc(0,-6){2}{Black}{Black}
\Text(60,0)[c]{$+~ \textnormal{(RSIF)} ~\times~$} 
\Text(83,-25)[c]{$(3)$}
\end{picture}
\hspace{1.2cm}
\begin{picture}(80,0)
\ArrowLine(0,0)(-15,-15)
\ArrowLine(-4,4)(-15,15)
\ArrowLine(15,15)(0,0)
\ArrowLine(15,-15)(5,-5)
\CCirc(0,0){2}{Black}{Black}
\CCirc(4,-4){2}{Black}{Black} 
\end{picture}\vspace{1.5cm}
$$
with the explicit expression\footnote{See Arxiv v4., of Ref.\cite{Dedes:2017zog} for the full (unsimplified) form.},
\begin{align*}
i\gam_{1234} &=   i\gam^{(1)}_{1234} + i\gam^{(2)}_{1234} + i\gam^{(3)}_{1234} \equiv i\gam^{(1,2)}_{1234}  + i\gam^{(3)}_{1234} ~,\\
i\gam^{(1)}_{1234} & = 2i \Big(C^{f_1f_2f_3f_4} (\gamma_\mu P_L)_{s_1s_2} (\gamma^\mu P_L)_{s_3s_4} - C^{f_2f_1f_3f_4} (\gamma_\mu P_R)_{s_1s_2} (\gamma^\mu P_L)_{s_3s_4} \\ &~~~~\,- C^{f_1f_2f_4f_3} (\gamma_\mu P_L)_{s_1s_2} (\gamma^\mu P_R)_{s_3s_4} + C^{f_2f_1f_4f_3} (\gamma_\mu P_R)_{s_1s_2} (\gamma^\mu P_R)_{s_3s_4}\Big)~,\\
i\gam^{(2)}_{1234} & = -i\gam^{(1)}_{1432}~,\\
i\gam^{(3)}_{1234} & = -i\gam^{(1)}_{1324}~,
\end{align*}
after applying the cc-operation and the Majorana condition. The last arrow-pair topology (\ie $\sim i\gam^{(3)}$) is particular to this four-Majorana vertex and creates the problem. As for B-violating, this issue is better understood from  the arrow-pair topologies which cannot be merged without clashing arrows or equivalently from the matrix elements, 
$$i\gam^{(1,2)}_{1234} \bar v_1 u_2 \bar v_3 u_4 ~,~~i\gam^{(3)}_{1234} \bar v_1 u_3 \bar v_2 u_4~, $$  
which cannot be merged without $\cc$ explicit.

One can deal in two ways with this problem. The first is to treat this vertex (only) with the traditional EFT approach. For that, the partially symmetrized quadrilinear,
\\[4mm]
$$~~~~\begin{picture}(60,30)
\Text(-25,-25)[c]{$\nu_1$}
\Text(-25,25)[c]{$\nu_2$}
\Text(25,25)[c]{$\nu_3$}
\Text(25,-25)[c]{$\nu_4$}
\ArrowArcn(-26,0)(20,0,-70)
\ArrowArcn(-26,0)(20,70, 0)
\ArrowArcn(26,0)(20,250,180)
\ArrowArcn(26,0)(20,180,110)
\CCirc(-6,0){2}{Black}{Black}
\CCirc(6,0){2}{Black}{Black}
\end{picture}
\longleftrightarrow~~~~~~~~~ {1\over 8} i \mathbb{\Gamma}^{(1)}_{1'2'3'4'}(\bar \nu_{1'}{\nu}_{2'}) (\bar{\nu}_{3'}{\nu_{4'}})~,
$$
\\[8mm]
is a useful shortcut as compared to the initial Lagrangian coupling. Interchange of contractions within the bilinears and the interchange of bilinears are equal here, canceling the symmetry factor. The other possible contractions are not, thus creating the two remaining topologies and the RSIFs. Being a partially symmetrized coupling, the naive real-scalar combinatorics will not work in the relevant graphs. Signs and combinatorics should be carefully watched but other than that, it is a straightforward procedure.

The other way is to use the full vertex by deviating from the algorithm. The fully-symmetrized quadrilinear participating in the Dyson expansion is now,
\\[4mm]
$$~~~~~\begin{picture}(0,30)
\Line(-20,-20)(20,20)
\Line(20,-20)(-20,20)
\CCirc(0,0){2}{Black}{Black}
\Text(-25,25)[c]{$\nu_2$}
\Text(25,-25)[c]{$\nu_4$}
\Text(25,25)[c]{$~\nu_3$}
\Text(-25,-25)[c]{$\nu_1$}
\end{picture}~~~~~~~~~~~~~\longleftrightarrow ~~~~~~{1\over 4!} \Big[(i \gam_{1'2'3'4'}^{(1)} + i \gam_{1'2'3'4'}^{(2)})( \bar\nu_{1'} \nu_{2'} \bar \nu_{3'}\nu_{4'}) + i \gam_{1'2'3'4'}^{(3)} ( \bar\nu_{1'} \bar \nu_{2'} \nu_{3'}\nu_{4'}) \Big]~,$$
\\[8mm]
so that all possible contractions are equal to each other and cancel the symmetry factor  for the vertex (\ie $S=4!$) even though they produce different Dirac-spinor structures. This form can be useful for code-implementation.

The diagrammatic approach is difficult but there can be various paths and shortcuts. Start with the diagram without clashing arrows and use only the primary vertex topology ($\sim i\gam^{(1,2)}$) as, 
\begin{align*}
\begin{picture}(40,40)
\LongArrow(-0.5,19.3)(-1.5,19.3)
\LongArrow(-15,-12)(-23,-12)
\LongArrow(23,-12)(15,-12)
\LongArrow(-14,-17)(-15,-17)
\LongArrow(15,-17)(14,-17)
\CArc(9,3)(20,125,240)
\CArc(10,3)(20,250,270)
\CArc(-10,3)(20,-90,55)
\Line(30,-17)(10.,-17)
\Line(-30,-17)(-10.,-17)
\CCirc(0,-14){3}{Black}{Black}
\Text(0,25)[c]{$\nu$}
\Text(-30,-25)[c]{$\nu_1$}
\Text(30,-25)[c]{$\nu_4$}
\Text(150,0)[c]{$\longrightarrow~~~i\mathcal M^{(1,2)} \sim + {i^2\over 2}\gam^{(1,2)}_{12'3'4} \bar u_1 S_{2'3'} u_4~. $}
\end{picture} ~~~~~~~~~~~~~~~~~~~~~~~~~~~~~~~~~~~~~~~~~~~~~~~
\\[1cm] 
\end{align*}
Next, replace \textit{iteratively} each primary vertex with a secondary (unattached) vertex ($\sim i\gam^{(3)}$), until only secondary vertices appear in the graph. Performing manipulations as, 
\begin{align*}
\begin{picture}(40,40)
\LongArrow(-0.5,19.3)(0.5,19.3)
\LongArrow(-15,-12)(-23,-12)
\LongArrow(23,-12)(15,-12)
\LongArrow(-14,-17)(-15,-17)
\LongArrow(15,-17)(14,-17)
\CArc(9,3)(20,125,220)
\CArc(10,3)(20,250,270)
\CArc(-10,3)(20,-90,55)
\Line(30,-17)(10.,-17)
\Line(-30,-17)(-10.,-17)
\CCirc(-5,-10){2}{Black}{Black}
\CCirc(0,-14){2}{Black}{Black}
\Text(0,25)[c]{$\nu$}
\Text(-30,-25)[c]{$\nu_1$}
\Text(30,-25)[c]{$\nu_4$}
\Text(50,0)[c]{$\longrightarrow$}
\end{picture} 
\hspace{2cm}
\begin{picture}(40,40)
\LongArrow(-15,-12)(-23,-12)
\LongArrow(23,-12)(15,-12)
\LongArrow(-14,-17)(-15,-17)
\LongArrow(15,-17)(14,-17)
\LongArrow(0.5,19.3)(-2,19.3)
\CArc(7,1)(20,115,230)
\CArc(10,3)(20,250,270)
\CArc(-10,3)(20,-90,-70)
\CArc(-7,1)(20,-60,70)
\Line(30,-17)(10.,-17)
\Line(-30,-17)(-10.,-17)
\CCirc(-3.5,-15.5){2}{Black}{Black}
\CCirc(3.5,-15.5){2}{Black}{Black}
\Text(0,25)[c]{$\nu$}
\Text(-30,-25)[c]{$\nu_1$}
\Text(30,-25)[c]{$\nu_4$}
\Text(120,0)[c]{$\longrightarrow ~~~i\mathcal M^{(3)} \sim + {i^2\over 2}\gam^{(1)}_{12'3'4} \bar u_1 S_{2'3'} u_4~,$}
\end{picture}\hspace{5cm}
\\[1cm] 
\end{align*}
the two-point function reads,
\begin{align*}
i\mathcal M = {i^2\over 2} \int {d^d k\over (2 \pi)^d}  ~(2\gam_{12'3'4}^{(1)} + \gam_{12'3'4}^{(2)}) \bar u_1 S_{2'3'}(k)   u_4~,
\end{align*}
with $S=2$ in the denominator being the overall symmetry factor, always following naive real-scalar combinatorics here.

For more involved graphs and four-point functions or higher, clashing arrows can appear when substituting primary with secondary vertices. By decomposing into arrow pair topologies they can be absorbed directly or through the external legs, as described in Sec. \ref{sec:fflow}. In general, one should proceed with caution when this vertex is relevant, keeping the Wick-contraction picture at hand.

%%%%%%%%%%% BIBLIOGRAPHY %%%%%%%%%%%%%%%%%
\bibliography{EFT}{}

\providecommand{\href}[2]{#2}\begingroup\raggedright\begin{thebibliography}{10}

\bibitem{Denner:1992vza}
A.~Denner, H.~Eck, O.~Hahn, and J.~Kublbeck, {\it {Feynman rules for fermion
  number violating interactions}},  {\em Nucl. Phys.} {\bf B387} (1992)
  467--481.

\bibitem{Denner:1992me}
A.~Denner, H.~Eck, O.~Hahn, and J.~Kublbeck, {\it {Compact Feynman rules for
  Majorana fermions}},  {\em Phys. Lett.} {\bf B291} (1992) 278--280.

\bibitem{Buras:1998raa}
A.~J. Buras, {\it {Weak Hamiltonian, CP violation and rare decays}},  in {\em
  {Probing the standard model of particle interactions. Proceedings, Summer
  School in Theoretical Physics, NATO Advanced Study Institute, 68th session,
  Les Houches, France, July 28-September 5, 1997. Pt. 1, 2}}, pp.~281--539,
  1998.
\newblock \href{http://arxiv.org/abs/hep-ph/9806471}{{\tt hep-ph/9806471}}.

\bibitem{Appelquist:1974tg}
T.~Appelquist and J.~Carazzone, {\it {Infrared Singularities and Massive
  Fields}},  {\em Phys. Rev.} {\bf D11} (1975) 2856.

\bibitem{Grzadkowski:2010es}
B.~Grzadkowski, M.~Iskrzynski, M.~Misiak, and J.~Rosiek, {\it {Dimension-Six
  Terms in the Standard Model Lagrangian}},  {\em JHEP} {\bf 10} (2010) 085,
  [\href{http://arxiv.org/abs/1008.4884}{{\tt arXiv:1008.4884}}].

\bibitem{Aebischer:2017ugx}
J.~Aebischer et~al., {\it {WCxf: an exchange format for Wilson coefficients
  beyond the Standard Model}},  \href{http://arxiv.org/abs/1712.05298}{{\tt
  arXiv:1712.05298}}.

\bibitem{Brivio:2017btx}
I.~Brivio, Y.~Jiang, and M.~Trott, {\it {The SMEFTsim package, theory and
  tools}},  {\em JHEP} {\bf 12} (2017) 070,
  [\href{http://arxiv.org/abs/1709.06492}{{\tt arXiv:1709.06492}}].

\bibitem{Celis:2017hod}
A.~Celis, J.~Fuentes-Martin, A.~Vicente, and J.~Virto, {\it {DsixTools: The
  Standard Model Effective Field Theory Toolkit}},
  \href{http://arxiv.org/abs/1704.04504}{{\tt arXiv:1704.04504}}.

\bibitem{Falkowski:2015wza}
A.~Falkowski, B.~Fuks, K.~Mawatari, K.~Mimasu, F.~Riva, and V.~sanz, {\it
  {Rosetta: an operator basis translator for Standard Model effective field
  theory}},  {\em Eur. Phys. J.} {\bf C75} (2015), no.~12 583,
  [\href{http://arxiv.org/abs/1508.05895}{{\tt arXiv:1508.05895}}].

\bibitem{Alonso:2013hga}
R.~Alonso, E.~E. Jenkins, A.~V. Manohar, and M.~Trott, {\it {Renormalization
  Group Evolution of the Standard Model Dimension Six Operators III: Gauge
  Coupling Dependence and Phenomenology}},  {\em JHEP} {\bf 04} (2014) 159,
  [\href{http://arxiv.org/abs/1312.2014}{{\tt arXiv:1312.2014}}].

\bibitem{Dedes:2017zog}
A.~Dedes, W.~Materkowska, M.~Paraskevas, J.~Rosiek, and K.~Suxho, {\it {Feynman
  rules for the Standard Model Effective Field Theory in R$_\xi$ -gauges}},
  {\em JHEP} {\bf 06} (2017) 143, [\href{http://arxiv.org/abs/1704.03888}{{\tt
  arXiv:1704.03888}}].

\bibitem{Brivio:2017vri}
I.~Brivio and M.~Trott, {\it {The Standard Model as an Effective Field
  Theory}},  \href{http://arxiv.org/abs/1706.08945}{{\tt arXiv:1706.08945}}.

\bibitem{Brivio:2016fzo}
I.~Brivio, J.~Gonzalez-Fraile, M.~C. Gonzalez-Garcia, and L.~Merlo, {\it {The
  complete HEFT Lagrangian after the LHC Run I}},  {\em Eur. Phys. J.} {\bf
  C76} (2016), no.~7 416, [\href{http://arxiv.org/abs/1604.06801}{{\tt
  arXiv:1604.06801}}].

\bibitem{Alanne:2017oqj}
T.~Alanne and F.~Goertz, {\it {Extended Dark Matter EFT}},
  \href{http://arxiv.org/abs/1712.07626}{{\tt arXiv:1712.07626}}.

\bibitem{Crivellin:2016ihg}
A.~Crivellin, M.~Ghezzi, and M.~Procura, {\it {Effective Field Theory with Two
  Higgs Doublets}},  {\em JHEP} {\bf 09} (2016) 160,
  [\href{http://arxiv.org/abs/1608.00975}{{\tt arXiv:1608.00975}}].

\bibitem{Jenkins:2017dyc}
E.~E. Jenkins, A.~V. Manohar, and P.~Stoffer, {\it {Low-Energy Effective Field
  Theory below the Electroweak Scale: Anomalous Dimensions}},  {\em JHEP} {\bf
  01} (2018) 084, [\href{http://arxiv.org/abs/1711.05270}{{\tt
  arXiv:1711.05270}}].

\bibitem{Jenkins:2017jig}
E.~E. Jenkins, A.~V. Manohar, and P.~Stoffer, {\it {Low-Energy Effective Field
  Theory below the Electroweak Scale: Operators and Matching}},
  \href{http://arxiv.org/abs/1709.04486}{{\tt arXiv:1709.04486}}.

\bibitem{Peskin:1995ev}
M.~E. Peskin and D.~V. Schroeder, {\em {An Introduction to quantum field
  theory}}.
\newblock Addison-Wesley, Reading, USA, 1995.

\end{thebibliography}\endgroup
\bibliographystyle{JHEP}                        
%%%%%%%%%%%%%%%%%%%%%%%%%%%%%%%%%%%

\end{document}